\documentclass[preprint,prd,aps,showpacs,showkeys,nofootinbib]{revtex4}
\usepackage{graphicx}
\begin{document}

\preprint{arXiv: 1706.08201}

\title{Evaluating Feynman integrals by the hypergeometry}

\author{Tai-Fu Feng$^{a,b}$\footnote{email:fengtf@hbu.edu.cn},
Chao-Hsi Chang$^{b,c,d}$\footnote{email:zhangzx@itp.ac.cn},
Jian-Bin Chen$^e$, Zhi-Hua Gu$^a$, Hai-Bin Zhang$^a$\footnote{email:hbzhang@hbu.edu.cn}}

\affiliation{$^a$Department of Physics, Hebei University, Baoding, 071002, China}
\affiliation{$^b$Key Laboratory of Theoretical Physics, Institute of Theoretical Physics,
Chinese Academy of Science, Beijing, 100190, China}
\affiliation{$^c$CCAST (World Laboratory), P.O.Box 8730, Beijing, 100190, China}
\affiliation{$^d$ School of Physical Sciences, University of Chinese Academy of Sciences, Beijing 100049, China}
\affiliation{$^e$Department of Physics, Taiyuan University of Technology, Taiyuan, 030024, China}

\begin{abstract}
The hypergeometric function method naturally provides the analytic expressions
of scalar integrals from concerned Feynman diagrams in some connected regions
of independent kinematic variables, also presents the systems of homogeneous linear partial
differential equations satisfied by the corresponding scalar integrals.
Taking examples of the one-loop $B_{_0}$ and massless $C_{_0}$ functions,
as well as the scalar integrals of two-loop vacuum and sunset diagrams,
we verify our expressions coinciding with the well-known results of literatures.
Based on the multiple hypergeometric functions of independent kinematic variables, the
systems of homogeneous linear partial differential equations satisfied by the mentioned scalar integrals
are established. Using the calculus of variations, one recognizes the system of
linear partial differential equations as stationary conditions of a functional under some
given restrictions, which is the cornerstone to perform the continuation of
the scalar integrals to whole kinematic domains numerically
with the finite element methods. In principle this method can be used to evaluate
the scalar integrals of any Feynman diagrams.
\end{abstract}


\maketitle

\section{Introduction\label{sec1}}
\indent\indent
The discovery of the Higgs particle in the Large Hadron Collider (LHC) implies that searching for
particle spectrum predicted by the standard model (SM) is finished now~\cite{CMS,ATLAS}.
One of the targets for particle physics now is to test the SM precisely and to search for
the new physics (NP) beyond the SM. Nevertheless
the experimental data from the running LHC seemingly
indicate the energy scale of new physics beyond the SM surpassing
$1$ TeV~\cite{PDG}. Thus the relevant corrections to the electroweak observables due to
new physics must be below 1\%. In order to discriminate the hints due to the new physics,
the SM backgrounds including two-loop electroweak corrections and multi-loop QCD
(quantum chromodynamics) corrections should be evaluated accurately.

The general dimensionally regularized scalar integrals can be expressed
as a linear combination of master integrals (irreducible scalar integrals)
through the method of integration by part (IBP)~\cite{K.G.Chetyrkin81}
for given Feynman diagrams. How to evaluate the scalar integrals
exactly is an obstacle to predict those electroweak observables precisely
in the SM. So far those one-loop scalar
integrals are calculated totally~\cite{G.'t.Hooft79,A.Denner11},
nevertheless the calculations of the multi-loop scalar integrals
are less advanced. The author of literature~\cite{V.A.Smirnov12} presents several methods
to evaluate those scalar integrals. Using Feynman parameterization method,
the author of Ref.~\cite{R.J.Gonsalves83} presents the analytic
expressions of the planar massless two-loop vertex diagrams.
The Mellin-Barnes (MB) method is often adopted to calculate some massless scalar
integrals~\cite{V.A.Smirnov99,J.B.Tausk99}, although the technique of multiple
MB representations is not optimal sometimes. Applying the IBP relations,
the authors of Refs.~\cite{A.V.Kotikov91-1,A.V.Kotikov91-2,A.V.Kotikov91-3,A.V.Kotikov91-4,
A.V.Kotikov92-1,S.Laporta96,S.Laporta97,E.Remiddi97,S.Laporta00,K.Melnikov00-1,K.Melnikov00-2,
M.Y.Kalmykov10,M.Y.Kalmykov12,M.Y.Kalmykov13,M.Y.Kalmykov17}
derive the differential equations on the master integrals
for a given set of Feynman integrals where the analytic expressions of
corresponding boundary conditions and inhomogeneous terms
are already known. Another method named 'dimensional recurrence
and analyticity' is also proposed by Refs.~\cite{R.N.Lee10-1,R.N.Lee10-2,R.N.Lee11-1,R.N.Lee11-2,
R.N.Lee11-3,R.N.Lee12-1,R.N.Lee12-2} to analyze the master integrals.
When a concerned scalar integral depends on kinematic invariants
and masses which essentially differ in order, the scalar integral
can be approached by the asymptotic expansions of momenta and
masses~\cite{V.A.Smirnov02}. In addition, various sector decompositions~\cite{K.Hepp66,E.R.Speer77}
are applied to numerically analyze the Feynman integrals~\cite{T.Kaneko10}.

Each method mentioned above has its blemishes, which can only be applied
to the Feynman diagrams with special topology and kinematic invariants.
Taking examples of the one-loop $B_{_0}$ and massless $C_{_0}$ functions,
as well as the scalar integrals from two-loop vacuum and sunset diagrams,
we here elucidate how to obtain the multiple hypergeometric functions
of independent kinematic variables for scalar integrals, which is convergent
in a connected region. Then we write down the
systems of homogeneous linear partial differential equations (PDEs) satisfied by
the corresponding multiple hypergeometric functions.
Generally the method provides different multiple hypergeometric
functions in the different kinematic domains for a given scalar integral,
certainly there is a system of homogeneous linear PDEs corresponding to each
multiple hypergeometric function. Nevertheless we can check directly
that those systems of homogeneous linear PDEs from different kinematic
regions are congruent with each other.
Actually the system of homogeneous linear PDEs
can be recognized as the stationary condition of a functional according to Hamilton's
principle~\cite{M.E.Taylor12}, thus the continuation of the scalar
integral in certain connected regions to whole domain of the kinematic invariants
and virtual masses is made numerically through the finite element methods.

The point specified here is that the system of homogeneous linear PDEs
derived from hypergeometric functions differs
from that presented in literatures ~\cite{A.V.Kotikov91-1,A.V.Kotikov91-2,
A.V.Kotikov91-3,A.V.Kotikov91-4,A.V.Kotikov92-1,S.Laporta96,S.Laporta97,
E.Remiddi97,S.Laporta00,K.Melnikov00-1,K.Melnikov00-2} obviously.
\begin{itemize}
\item Here the system of homogeneous linear PDEs is
derived from the convergent multiple hypergeometric functions in the connected regions,
while the PDEs of literatures are based on the IBP relations.
\item The concerned scalar integral is the only unknown function to be determined in
our system of PDEs.  Correspondingly several
master integrals originating from a given set of Feynman diagrams are generally
coupled in the PDEs from the literatures.
\item The PDEs in our approach are linear and homogeneous. While
the PDEs in the literatures above generally contain
inhomogeneous terms which are known already.
\item Using the convergent multiple hypergeometric functions in certain
connected regions, we perform the continuation of the scalar integral
numerically through the system of PDEs here. In literatures above, one determines
the solutions to the PDEs using the boundary conditions whose
analytic expressions are known also.
\item In addition our systems of linear PDEs embody
the possible interchanging symmetries among the independent variables explicitly.
\end{itemize}

The method to derive the multiple hypergeometric functions here,
also named $x-$space technique in literature, is discussed in
Refs.~\cite{E.Mendels78,F.A.Berends94}. Ignoring all virtual masses, the authors of
Ref.~\cite{K.G.Chetyrkin83} apply this method to derive the renormalization group equations (RGEs),
and the authors of Ref.~\cite{K.G.Chetyrkin80} analyze three-loop ratio $R(s)$ in QCD.
Here the equivalency between the traditional Feynman parameterization and the hypergeometric
function method can be proved by
the integral representation of Bessel functions~\cite{G.N.Watson44}.
Applying theory of generalized hypergeometric function~\cite{L.J.Slater66}, we
present the multiple hypergeometric functions for the one-loop $B_{_0}$
function, two-loop vacuum integral, and the scalar integrals from two-loop
sunset and one-loop 3-point diagrams, respectively.
Those multiple hypergeometric functions are convergent
in some connected regions of the independent kinematic variables.
Thus the systems of homogeneous linear PDEs satisfied by the corresponding
multiple hypergeometric series are established explicitly.
With the systems of homogeneous linear PDEs, one may numerically evaluate the
necessary values correctly.

Our presentation is organized as follows. Taking example of $B_{_0}$ function,
we prove firstly the equivalency between the traditional Feynman parameterization
and the hypergeometric function method in section \ref{sec3}.
Then we present the convergent double hypergeometric series of
the one-loop $B_{_0}$ function of certain connected regions together with
the system of homogeneous linear PDEs describing the properties of the one-loop
$B_{_0}$ function in whole domain of independent kinematic variables.
Using some well-known reduction formulae of Apell functions, one recovers the
expression of the one-loop $B_{_0}$ function from textbook~\cite{A.Denner93}
explicitly in section \ref{sec3} also. The similar convergent
multiple hypergeometric functions of the two-loop vacuum scalar integral and the corresponding systems
of homogeneous linear PDEs are presented in section \ref{sec4}, and that of
the scalar integral from two-loop sunset diagram are given in section \ref{sec5}, separately.
The corresponding systems of homogeneous linear PDEs
for the scalar integral of massless one-loop triangle diagram are given in
the section \ref{sec6}, meanwhile the comparison of our expression with
the well-known result of literature is also presented briefly.
In the section \ref{sec7}, we recognize the systems of linear PDEs
as the stationary conditions of a functional under some restrictions
according to Hamilton's principle, which is convenient for numerically
evaluating the scalar integrals through the finite element method. Finally our
conclusions are summarized in section \ref{sec8}.

\section{The system of homogeneous linear PDEs for $B_{_0}$ function\label{sec3}}
\indent\indent
In the $D-$dimension Euclidean space, the modified Bessel functions can be
written as~\cite{G.N.Watson44}
\begin{eqnarray}
&&{2(m^2)^{D/2-\alpha}\over(4\pi)^{D/2}\Gamma(\alpha)}
k_{_{D/2-\alpha}}(mx)
=\int{d^Dq\over(2\pi)^D}{\exp[-i{\bf q}\cdot {\bf x}]\over(q^2+m^2)^\alpha}
\;,\nonumber\\
&&{\Gamma(D/2-\alpha)\over(4\pi)^{D/2}\Gamma(\alpha)}\Big({x\over2}\Big)^{2\alpha-D}
=\int{d^Dq\over(2\pi)^D}{\exp[-i{\bf q}\cdot {\bf x}]\over(q^2)^\alpha}
\;,\nonumber\\
&&2\pi^{D/2}j_{_{D/2-1}}(qx)=\int d^{D-1}\hat{\bf x}\exp[i{\bf q}\cdot {\bf x}]\;,
\label{base1}
\end{eqnarray}
where ${\bf q}$, ${\bf x}$ are vectors, and $d^{D-1}\hat{\bf x}$ denotes angle integral,
respectively. With those identities, the one-loop $B_{_0}$ function is formulated as
\begin{eqnarray}
&&B_{_0}(p^2)=
{i4(m_{_1}^2m_{_2}^2)^{D/2-1}(\mu^2)^{2-D/2}\over(4\pi)^{D/2}}\int dx
\Big({x\over2}\Big)^{D-1}j_{_{D/2-1}}(p_{_{\rm E}}x)k_{_{D/2-1}}(m_{_1}x)k_{_{D/2-1}}(m_{_2}x)\;,
\label{1LoopB0-1}
\end{eqnarray}
where ${\bf p}_{_{\rm E}}$ represents the momentum in the Euclidean space,
$p_{_{\rm E}}=|{\bf p}_{_{\rm E}}|$,
and $\mu$ denotes the renormalization energy scale, respectively.
In order to verify the equivalency between Feynman parameterization
and the hypergeometric function method, one applies the integral representation
of the Bessel function
\begin{eqnarray}
&&k_{_\mu}(x)={1\over2}\int_0^\infty t^{-\mu-1}\exp\{-t-{x^2\over4t}\}dt\;,
\;\;\;\Re(x^2)>0\;.
\label{base18}
\end{eqnarray}
Thus the one-loop $B_{_0}$ function is written as
\begin{eqnarray}
&&B_{_0}(p^2)=
{i(\mu^2)^{2-D/2}\over(4\pi)^{D/2}}\int_0^\infty dt_{_1}\int_0^\infty dt_{_2}
{\exp\Big\{-m_{_1}^2t_{_1}-m_{_2}^2t_{_2}-{t_{_1}t_{_2}p_{_{\rm E}}^2\over t_{_1}+t_{_2}}\Big\}
\over(t_{_1}+t_{_2})^{D/2}}\;.
\label{1LoopB0-54}
\end{eqnarray}
Performing the variable transformation
\begin{eqnarray}
&&t_{_1}=\varrho(1-y),\;t_{_2}=\varrho y\;,
\label{1LoopB0-55a}
\end{eqnarray}
where the Jacobi of the transformation is
\begin{eqnarray}
&&{\partial(t_{_1},t_{_2})\over\partial(\varrho,y)}=\varrho\;,
\label{1LoopB0-55}
\end{eqnarray}
we finally have
\begin{eqnarray}
&&B_{_0}(p^2)=
{i(\mu^2)^{2-D/2}\over(4\pi)^{D/2}}\int_0^1 dy\int_0^\infty d\varrho\varrho^{1-D/2}
\exp\Big\{-\varrho\Big(m_{_1}^2y+m_{_2}^2(1-y)+y(1-y)p_{_{\rm E}}^2\Big)\Big\}
\nonumber\\
&&\hspace{1.3cm}=
{i(\mu^2)^{2-D/2}\Gamma(2-{D\over2})\over(4\pi)^{D/2}}
\int_0^1 dy{1\over\Big(m_{_1}^2y+m_{_2}^2(1-y)+y(1-y)p_{_{\rm E}}^2\Big)^{2-{D\over2}}}\;.
\label{1LoopB0-56}
\end{eqnarray}
Replacing the momentum squared of Euclidean space $p_{_{\rm E}}^2$ with
that of Minkowski space $-p^2$, one finds that the expression of
$B_{_0}$ function in Eq.(\ref{1LoopB0-56}) can be recovered
from the Feynman parameterization exactly.

In order to obtain the double hypergeometric series for one-loop $B_{_0}$ function,
we present the power series of modified Bessel functions as
\begin{eqnarray}
&&j_{_{\mu}}(x)=\sum\limits_{n=0}^\infty{(-1)^n\over n!\Gamma(1+\mu+n)}\Big({x\over2}\Big)^{2n}\;,\nonumber\\
&&k_{_{\mu}}(x)={\Gamma(\mu)\Gamma(1-\mu)\over2}\sum\limits_{n=0}^\infty{1\over n!}\Big[
-{1\over\Gamma(1+\mu+n)}\Big({x\over2}\Big)^{2n}+{1\over\Gamma(1-\mu+n)}\Big({x\over2}\Big)^{2(n-\mu)}\Big]\;.
\label{base6}
\end{eqnarray}
Inserting the expressions of $k_{_{D/2-1}}(m_{_1}x),\;k_{_{D/2-1}}(m_{_2}x)$
into Eq.(\ref{1LoopB0-1}) and applying the radial integral 
\begin{eqnarray}
&&\int_0^\infty dt \Big({t\over2}\Big)^{2\varrho-1}k_{_\mu}(t)={1\over2}
\Gamma(\varrho)\Gamma(\varrho-\mu)\;,
\nonumber\\
&&\int_0^\infty dt \Big({t\over2}\Big)^{2\varrho-1}j_{_\mu}(t)=
{\Gamma(\varrho)\over\Gamma(1-\varrho+\mu)}
\label{base15}
\end{eqnarray}
as $|p^2|>\max(m_{_1}^2,m_{_2}^2)$, 
one writes the analytic expression of the $B_{_0}$ function as
\begin{eqnarray}
&&B_{_0}(p^2)={i(-p^2)^{D/2-2}\over(4\pi)^{D/2}(\mu^2)^{D/2-2}}
{\Gamma(3-D/2)\over D-3}\varphi_{_1}(x,y)\;,
\label{1LoopB0-3a}
\end{eqnarray}
with $x=m_{_1}^2/p^2,\;y=m_{_2}^2/p^2$. Meanwhile the double
hypergeometric functions $\varphi_{_1}(x,y)$ is
\begin{eqnarray}
&&\varphi_{_1}(x,y)=
{D-3\over({D\over2}-2)({D\over2}-1)}\Big\{\Big(-x\Big)^{D/2-1}F_{_4}
\left(\left.\begin{array}{cc}1,&2-{D\over2}\\
{D\over2},&2-{D\over2}\end{array}\right|x,\;y\right)
\nonumber\\
&&\hspace{1.9cm}
+\Big(-y\Big)^{D/2-1}F_{_4}
\left(\left.\begin{array}{cc}1,&2-{D\over2}\\
2-{D\over2},&{D\over2}\end{array}\right|x,\;y\right)
\nonumber\\
&&\hspace{1.9cm}
-{\Gamma(D/2)\Gamma(D/2-1)\over\Gamma(D-2)}F_{_4}
\left(\left.\begin{array}{cc}2-{D\over2},&3-D\\
2-{D\over2},&2-{D\over2}\end{array}\right|x,\;y\right)
\Big\}\;,
\label{1LoopB0-3}
\end{eqnarray}
where $F_{_4}$ is the Apell function
\begin{eqnarray}
&&F_{_4}\left(\left.\begin{array}{cc}a,&b\\
c_{_1},&c_{_2}\end{array}\right|x,\;y\right)
=\sum\limits_{m=0}^\infty\sum\limits_{n=0}^\infty{(a)_{_{m+n}}(b)_{_{m+n}}
\over m!n!(c_{_1})_{_m}(c_{_2})_{_n}}x^my^n
\label{1LoopB0-4}
\end{eqnarray}
whose convergent region is $\sqrt{|x|}+\sqrt{|y|}\le1$. Here we adopt
the abbreviation used in Ref.~\cite{L.J.Slater66}
\begin{eqnarray}
&&(a)_{_m}={\Gamma(a+m)\over\Gamma(a)}\;.
\label{1LoopB0-4-1}
\end{eqnarray}
Obviously the double hypergeometric function $\varphi_{_1}(x,y)$ satisfies
the system of homogeneous linear PDEs
\begin{eqnarray}
&&\Big\{(\hat{\vartheta}_{x}+\hat{\vartheta}_{y}+2-{D\over2})(\hat{\vartheta}_{x}
+\hat{\vartheta}_{y}+3-D)-{1\over x}\hat{\vartheta}_{x}(\hat{\vartheta}_{x}+1
-{D\over2})\Big\}\varphi_{_1}=0
\;,\nonumber\\
&&\Big\{(\hat{\vartheta}_{x}+\hat{\vartheta}_{y}+2-{D\over2})(\hat{\vartheta}_{x}
+\hat{\vartheta}_{y}+3-D)-{1\over y}\hat{\vartheta}_{y}(\hat{\vartheta}_{y}+1
-{D\over2})\Big\}\varphi_{_1}=0\;,
\label{1LoopB0-5}
\end{eqnarray}
with $\hat{\vartheta}_{x}=x{\partial/\partial x}$.

Similarly inserting the power series of
$k_{_{D/2-1}}(m_{_1}x),\;j_{_{D/2-1}}(px)$ into Eq.(\ref{1LoopB0-1})
and applying radial integral in Eq.(\ref{base15}), we formulate the $B_{_0}$ function as
\begin{eqnarray}
&&B_{_0}(p^2)={i(m_{_2}^2)^{D/2-2}\over(4\pi)^{D/2}(\mu^2)^{D/2-2}}
{\Gamma(3-D/2)\over D-3}\varphi_{_2}(\xi,\eta)\;,
\label{1LoopB0-7a}
\end{eqnarray}
with $\xi=p^2/m_{_2}^2,\;\eta=m_{_1}^2/m_{_2}^2$. Furthermore,
the double hypergeometric function $\varphi_{_2}(\xi,\eta)$ is given as
\begin{eqnarray}
&&\varphi_{_2}(\xi,\eta)=
{D-3\over({D\over2}-2)({D\over2}-1)}
\Big\{\eta^{D/2-1}F_{_4}
\left(\left.\begin{array}{cc}1,&{D\over2}\\
{D\over2},&{D\over2}\end{array}\right|\xi,\;\eta\right)
\nonumber\\
&&\hspace{1.9cm}
-F_{_4}\left(\left.\begin{array}{cc}1,&2-{D\over2}\\
{D\over2},&2-{D\over2}\end{array}\right|\xi,\;\eta\right)\Big\}\;,
\label{1LoopB0-7}
\end{eqnarray}
whose convergent region is $\sqrt{|\xi|}+\sqrt{|\eta|}\le1$,
or equivalently $1+\sqrt{|x|}\le\sqrt{|y|}$.
Correspondingly the double hypergeometric function $\varphi_{_2}(\xi,\eta)$
satisfies the system of homogeneous linear PDEs
\begin{eqnarray}
&&\Big\{(\hat{\vartheta}_{\xi}+\hat{\vartheta}_{\eta}+1)
(\hat{\vartheta}_{\xi}+\hat{\vartheta}_{\eta}+2-{D\over2})
-{1\over\xi}\hat{\vartheta}_{\xi}(\hat{\vartheta}_{\xi}-1
+{D\over2})\Big\}\varphi_{_2}=0
\;,\nonumber\\
&&\Big\{(\hat{\vartheta}_{\xi}+\hat{\vartheta}_{\eta}+1)
(\hat{\vartheta}_{\xi}+\hat{\vartheta}_{\eta}+2-{D\over2})
-{1\over\eta}\hat{\vartheta}_{\eta}(\hat{\vartheta}_{\eta}+1
-{D\over2})\Big\}\varphi_{_2}=0\;.
\label{1LoopB0-9}
\end{eqnarray}
Interchanging $m_{_1}\leftrightarrow m_{_2}$ in the double hypergeometric
function of Eq.(\ref{1LoopB0-7a}) and the system of PDEs
of Eq.(\ref{1LoopB0-9}), one obtains the corresponding results of the case
$m_{_1}^2>\max(|p^2|,m_{_2}^2)$. A point specified here is that
two systems of homogeneous linear PDEs in
Eq.(\ref{1LoopB0-5}) and Eq.(\ref{1LoopB0-9}) are equivalent.
Inserting $\varphi_{_2}(\xi,\eta)=(-y)^{2-D/2}\varphi_{_1}(x,y),\;\xi=1/y,\;
\eta=x/y$ into Eq.(\ref{1LoopB0-9}), one derives two linear combinations of
the PDEs in Eq.(\ref{1LoopB0-7a})
directly. This implicates that the function defined through Eq.(\ref{1LoopB0-3a})
satisfies the system of homogeneous linear PDEs
of Eq.(\ref{1LoopB0-5}) outside the convergent region of the double hypergeometric
series in Eq.(\ref{1LoopB0-3}). In other words, the continuation of
$\varphi_{_1}$ from its convergent regions to the whole kinematic domain
can be achieved numerically through the system of homogeneous linear PDEs.
We will address this point in detail in section \ref{sec7}.

In order to recover the expression of the one-loop $B_{_0}$ function in textbook~\cite{A.Denner93},
we need the well-known reduction formulae~\cite{L.J.Slater66}
\begin{eqnarray}
&&F_{_4}\left(\left.\begin{array}{cc}\alpha,&\beta\\
\beta,&\beta\end{array}\right|-{u\over(1-u)(1-v)},\;-{v\over(1-u)(1-v)}\right)
\nonumber\\
&&\hspace{-0.5cm}=
(1-u)^\alpha(1-v)^\alpha\;_{_2}F_{_1}
\left(\left.\begin{array}{c}\alpha,\;1+\alpha-\beta\\
\beta\end{array}\right|uv\right)
\;,\nonumber\\
&&F_{_4}\left(\left.\begin{array}{c}\alpha,\;\beta\\
1+\alpha-\beta,\;\beta\end{array}\right|-{u\over(1-u)(1-v)},\;-{v\over(1-u)(1-v)}\right)
\nonumber\\
&&\hspace{-0.5cm}=
(1-v)^\alpha\;_{_2}F_{_1}
\left(\left.\begin{array}{c}\alpha,\;\beta\\
1+\alpha-\beta\end{array}\right|-{u(1-v)\over1-u}\right)
\;,\nonumber\\
&&\;_{_1}F_{_0}(\left.\alpha\;\right|x)=(1-x)^{-\alpha}\;.
\label{1LoopB0-46}
\end{eqnarray}
As $\max(|x|,|y|)\le1$ and $\lambda_{_{x,y}}^2=1+x^2+y^2-2x-2y-2xy\ge0$,
we find
\begin{eqnarray}
&&B_{_0}(p^2)=
{i\over(4\pi)^{D/2}}{\Gamma(3-D/2)\over D-3}\Big({-p^2\over\mu^2}\Big)^{D/2-2}
\varphi_{_1}(x,y)
\nonumber\\
&&\hspace{1.3cm}=
{i\Gamma(D/2-1)\Gamma(2-D/2)\over(4\pi)^{D/2}}
\Big({-p^2\over\mu^2}\Big)^{D/2-2}
\nonumber\\
&&\hspace{1.5cm}\times
\Big\{-{(-x)^{D/2-1}\over\Gamma(D/2)}(1-v)
\;_{_2}F_{_1}\left(\left.\begin{array}{c}1,\;2-D/2\\ D/2\end{array}\right|
-{u(1-v)\over1-u}\right)
\nonumber\\
&&\hspace{1.5cm}
-{(-y)^{D/2-1}\over\Gamma(D/2)}(1-u)
\;_{_2}F_{_1}\left(\left.\begin{array}{c}1,\;2-D/2\\ D/2\end{array}\right|
-{v(1-u)\over1-v}\right)
\nonumber\\
&&\hspace{1.5cm}
+{\Gamma(D/2-1)\over\Gamma(D-2)}\Big({1-uv\over(1-u)(1-v)}\Big)^{D-3}\Big\}\;,
\label{1LoopB0-47a}
\end{eqnarray}
with
\begin{eqnarray}
&&u={-1+x+y+\lambda_{_{x,y}}\over2y},\;v={-1+x+y+\lambda_{_{x,y}}\over2x}\;.
\label{1LoopB0-48}
\end{eqnarray}

Using $D=4-2\varepsilon$ and keeping terms up to ${\cal O}(\varepsilon^2)$ only,
one easily obtains the following expansion
\begin{eqnarray}
&&\;_{_2}F_{_1}\left(\left.\begin{array}{c}\varepsilon,\;1\\
2-\varepsilon\end{array}\right|x\right)
\simeq{1-\varepsilon\over1-2\varepsilon}\Big\{1+{1-x\over x}\Big[\varepsilon\ln(1-x)
\nonumber\\
&&\hspace{4.1cm}
-\varepsilon^2\Big(\ln^2(1-x)+L_{_{i_2}}(x)\Big)\Big]\Big\}\;,\nonumber\\
&&\Gamma(x+\varepsilon)=\Big[1+\varepsilon\psi(x)+{\varepsilon^2\over2}\Big(\psi^\prime(x)
+\psi^2(x)\Big)+\cdots\Big]\Gamma(x)\;.
\label{1LoopB0-46b}
\end{eqnarray}
With the expansions of Eq.(\ref{1LoopB0-46b}) and some well-known
identities of dilogarithm functions, the Laurent series of
the one-loop $B_{_0}$ function around $D=4$ is obtained as
\begin{eqnarray}
&&B_{_0}(p^2)\simeq
{i\Gamma(1+\varepsilon)\over(1-2\varepsilon)(4\pi)^2}
\Big({4\pi\mu^2\over-p^2}\Big)^\varepsilon
\Big\{{1\over\varepsilon}
+\Big[-{1\over2}\ln(xy)-{x-y\over2}\ln{x\over y}
\nonumber\\
&&\hspace{1.7cm}
-\lambda_{_{x,y}}\ln{1-x-y-\lambda_{_{x,y}}\over2\sqrt{xy}}\Big]
+\varepsilon\Phi_{_1}(x,y)+\cdots\Big\}\;,
\label{1LoopB0-47}
\end{eqnarray}
where the first two terms coincide with the well-known expression
of one-loop $B_{_0}$ function in text book~\cite{A.Denner93}.
The function $\Phi_{_1}(x,y)$ in this kinematic region is given as
\begin{eqnarray}
&&\Phi_{_1}(x,y)=-(1-\lambda_{_{x,y}}){\pi^2\over6}+{1+x-y-\lambda_{_{x,y}}\over2}
L_{_{i_2}}({\lambda_{_{x,y}}(1-x-y-\lambda_{_{x,y}})\over x(1-x+y-\lambda_{_{x,y}})})
\nonumber\\
&&\hspace{2.0cm}
+{1-x+y-\lambda_{_{x,y}}\over2}
L_{_{i_2}}({\lambda_{_{x,y}}(1-x-y-\lambda_{_{x,y}})\over y(1+x-y-\lambda_{_{x,y}})})
\nonumber\\
&&\hspace{2.0cm}
+\lambda_{_{x,y}}\ln{\lambda_{_{x,y}}(1-x-y-\lambda_{_{x,y}})
\over x(1-x+y-\lambda_{_{x,y}})}\ln(-x)
\nonumber\\
&&\hspace{2.0cm}
+\lambda_{_{x,y}}\ln{\lambda_{_{x,y}}(1-x-y-\lambda_{_{x,y}})
\over y(1+x-y-\lambda_{_{x,y}})}\ln(-y)
\nonumber\\
&&\hspace{2.0cm}
+{1+x-y-\lambda_{_{x,y}}\over4}\ln^2(-x)+{1-x+y-\lambda_{_{x,y}}\over4}\ln^2(-y)
\nonumber\\
&&\hspace{2.0cm}
+{1+x-y-\lambda_{_{x,y}}\over2}\ln{1+x-y-\lambda_{_{x,y}}\over
2\lambda_{_{x,y}}}\ln{\lambda_{_{x,y}}(1-x-y-\lambda_{_{x,y}})
\over x(1-x+y-\lambda_{_{x,y}})}
\nonumber\\
&&\hspace{2.0cm}
+{1-x+y-\lambda_{_{x,y}}\over2}\ln{1-x+y-\lambda_{_{x,y}}\over
2\lambda_{_{x,y}}}\ln{\lambda_{_{x,y}}(1-x-y-\lambda_{_{x,y}})
\over y(1+x-y-\lambda_{_{x,y}})}\;,
\label{1LoopB0-47b}
\end{eqnarray}
which can be used to extract the corrections from one-loop self-energy
counter term diagrams.

Using the quadratic transformation, one makes the analytic continuation of
the corresponding expression of the kinematic region $\lambda_{_{x,y}}^2\ge0$ 
to the region $\lambda_{_{x,y}}^2\le0$.
The useful quadratic transformations of Gauss functions are
\begin{eqnarray}
&&\;_{_2}F_{_1}\left(\left.\begin{array}{c}a,b\\ 1+a-b\end{array}\right|
\xi\right)=(1-\xi)^{-a}\;_{_2}F_{_1}\left(\left.\begin{array}{c}{a\over2},\;1/2+a/2-b\\ 1+a-b\end{array}\right|
-{4\xi\over(1-\xi)^2}\right)
\;,\nonumber\\
&&\;_{_2}F_{_1}\left(\left.\begin{array}{c}a,b\\ c\end{array}\right|
\xi\right)={\Gamma(c)\Gamma(b-a)\over\Gamma(b)\Gamma(c-a)}(-\xi)^{-a}
\;_{_2}F_{_1}\left(\left.\begin{array}{c}a,1+a-c\\ 1+a-b\end{array}\right|{1\over\xi}\right)
\nonumber\\
&&\hspace{3.2cm}
+{\Gamma(c)\Gamma(a-b)\over\Gamma(a)\Gamma(c-b)}(-\xi)^{-b}
\;_{_2}F_{_1}\left(\left.\begin{array}{c}b,1+b-c\\ 1+b-a
\end{array}\right|{1\over\xi}\right)\;,\nonumber\\
&&\;_{_2}F_{_1}\left(\left.\begin{array}{c}a,b\\ c\end{array}\right|
\xi\right)=(1-\xi)^{c-a-b}\;_{_2}F_{_1}\left(\left.\begin{array}{c}c-a,c-b\\ c\end{array}\right|
\xi\right)\;,
\label{1LoopB0-51}
\end{eqnarray}
if $|\arg(-\xi)|<\pi$.
Applying the quadratic transformation on the Gauss functions in Eq.(\ref{1LoopB0-47a}),
one has
\begin{eqnarray}
&&B_{_0}(p^2)=
{i\Gamma(1+\varepsilon)\over(4\pi)^2\varepsilon(1-2\varepsilon)}
\Big({4\pi\mu^2\over-p^2}\Big)^{\varepsilon}
\nonumber\\
&&\hspace{1.8cm}\times
\Big\{x^{1/2}(-x)^{-\varepsilon}
\Big(1+{\lambda_{_{x,y}}^2\over4x}\Big)^{1/2}
\;_{_2}F_{_1}\left(\left.\begin{array}{c}\varepsilon,1\\{1\over2}+\varepsilon\end{array}\right|
-{\lambda_{_{x,y}}^2\over4x}\right)
\nonumber\\
&&\hspace{1.8cm}
+y^{1/2}(-y)^{-\varepsilon}\Big(1+{\lambda_{_{x,y}}^2\over4y}\Big)^{1/2}
\;_{_2}F_{_1}\left(\left.\begin{array}{c}\varepsilon,1\\{1\over2}+\varepsilon\end{array}\right|
-{\lambda_{_{x,y}}^2\over4y}\right)
\nonumber\\
&&\hspace{1.8cm}
-{\Gamma(1-\varepsilon)\Gamma(\varepsilon+1/2)(-4)^{\varepsilon}\over\pi^{1/2}}
\lambda_{_{x,y}}^{1-2\varepsilon}
+{\Gamma^2(1-\varepsilon)\over\Gamma(1-2\varepsilon)}\lambda_{_{x,y}}^{1-2\varepsilon}\Big\}\;.
\label{1LoopB0-52a}
\end{eqnarray}
In order to continue our analyses, we adopt the following expansion and
reduction formulae~\cite{L.J.Slater66}
\begin{eqnarray}
&&\;_{_2}F_{_1}\left(\left.\begin{array}{c}\varepsilon,1\\{1\over2}+\varepsilon\end{array}\right|
t\right)\simeq1+2\varepsilon t\;_{_2}F_{_1}\left(\left.\begin{array}{c}1,1\\{3\over2}\end{array}\right|
t\right)
\nonumber\\
&&\hspace{-0.5cm}
+2\varepsilon^2t\Big[\partial_{_a}\;_{_2}F_{_1}\left(\left.\begin{array}{c}1,1\\{3\over2}\end{array}\right|
t\right)+\partial_{_c}\;_{_2}F_{_1}\left(\left.\begin{array}{c}1,1\\{3\over2}\end{array}\right|
t\right)-2\;_{_2}F_{_1}\left(\left.\begin{array}{c}1,1\\{3\over2}\end{array}\right|
t\right)\Big]
\;,\nonumber\\
&&\;_{_2}F_{_1}\left(\left.\begin{array}{c}1,1\\{3\over2}\end{array}\right|
t\right)={\arcsin\sqrt{t}\over\sqrt{t(1-t)}}
\;,\nonumber\\
&&\partial_{_a}\;_{_2}F_{_1}\left(\left.\begin{array}{c}1,1\\{3\over2}\end{array}\right|
t\right)+\partial_{_c}\;_{_2}F_{_1}\left(\left.\begin{array}{c}1,1\\{3\over2}\end{array}\right|
t\right)-2\;_{_2}F_{_1}\left(\left.\begin{array}{c}1,1\\{3\over2}\end{array}\right|t\right)
\nonumber\\
&&\hspace{-0.5cm}=
-{1\over\sqrt{t(1-t)}}\Big[\ln(4t)\arcsin\sqrt{t}+{\rm Cl}_{_2}(2\arcsin\sqrt{t})\Big]\;,
\label{1LoopB0-51b}
\end{eqnarray}
where ${\rm Cl}_{_2}$ denotes the Clausen function.
Thus the $B_{_0}$ function of the kinematic region
$\lambda_{_{x,y}}^2\le0$ is written as
\begin{eqnarray}
&&B_{_0}(p^2)\simeq
{i\Gamma(1+\varepsilon)\over(4\pi)^2(1-2\varepsilon)}\Big({4\pi\mu^2\over-p^2}\Big)^{\varepsilon}
\Big\{{1\over\varepsilon}
+\Big[-{1\over2}\ln(xy)-{x-y\over2}\ln{x\over y}
\nonumber\\
&&\hspace{1.8cm}
-\lambda_{_{x,y}}\ln{1-x-y-\lambda_{_{x,y}}\over2\sqrt{xy}}\Big]
+\varepsilon\Phi_{_1}(x,y)+\cdots\Big\}\;.
\label{1LoopB0-52}
\end{eqnarray}
Where the first two terms are consistent with the well-known results
of the one-loop $B_{_0}$ function, and
the function $\Phi_{_1}(x,y)$ in kinematic regions $\lambda_{_{x,y}}^2\le0$
is formulated as
\begin{eqnarray}
&&\Phi_{_1}(x,y)={1+x-y\over4}\ln^2(-x)+{1-x+y\over4}\ln^2(-y)
\nonumber\\
&&\hspace{2.0cm}
-\sqrt{-\lambda_{_{x,y}}^2}\ln(-\lambda_{_{x,y}}^2)[\arcsin({1+x-y\over2\sqrt{x}})
+\arcsin({1-x+y\over2\sqrt{y}})]
\nonumber\\
&&\hspace{2.0cm}
-\sqrt{-\lambda_{_{x,y}}^2}[{\rm Cl}_{_2}(2\arcsin({1+x-y\over2\sqrt{x}}))
+{\rm Cl}_{_2}(2\arcsin({1-x+y\over2\sqrt{y}}))]
\nonumber\\
&&\hspace{2.0cm}
+{\lambda_{_{x,y}}\over2}\Big([\psi(1)-\ln\lambda_{_{x,y}}^2]^2-4\psi^\prime(1)
-\psi^\prime(1/2)-\ln^2(-\lambda_{_{x,y}}^2)\Big)\;.
\label{1LoopB0-52b}
\end{eqnarray}

As $m_{_2}^2>\max(m_{_1}^2,\;|p^2|)$ and
$\lambda_{_{\xi,\eta}}^2=1+\xi^2+\eta^2-2\xi-2\eta-2\xi\eta\ge0$,
the $B_{_0}$ function is similarly written as
\begin{eqnarray}
&&B_{_0}(p^2)=
{i\over(4\pi)^{D/2}}{\Gamma(D/2-1)\Gamma(2-D/2)\over\Gamma(D/2)}
\Big({m_{_2}^2\over\mu^2}\Big)^{D/2-2}
\nonumber\\
&&\hspace{1.8cm}\times
\Big\{-\eta^{D/2-1}(1-z)(1-w)
\;_{_2}F_{_1}\left(\left.\begin{array}{c}1,2-{D\over2}\\ {D\over2}\end{array}\right|
zw\right)
\nonumber\\
&&\hspace{1.8cm}
+(1-w)\;_{_2}F_{_1}\left(\left.\begin{array}{c}1,2-{D\over2}\\{D\over2}\end{array}\right|
-{z(1-w)\over1-z}\right)\Big\}
\label{1LoopB0-49}
\end{eqnarray}
with
\begin{eqnarray}
&&z={-1+\xi+\eta+\lambda_{_{\xi,\eta}}\over2\eta},\;
w={-1+\xi+\eta+\lambda_{_{\xi,\eta}}\over2\xi}\;.
\label{1LoopB0-50}
\end{eqnarray}

Using the expansion of Eq.(\ref{1LoopB0-46b}), one finally gets
\begin{eqnarray}
&&B_{_0}(p^2)\simeq
{i\over(4\pi)^2}\Big({4\pi\mu^2\over m_{_2}^2}\Big)^\varepsilon
{\Gamma(1+\varepsilon)\over1-2\varepsilon}
\Big\{{1\over\varepsilon}
+\Big[-{\lambda_{_{\xi,\eta}}\over\xi}\ln{\lambda_{_{\xi,\eta}}
(1-\xi+\eta-\lambda_{_{\xi,\eta}})\over2\sqrt{\eta}}
\nonumber\\
&&\hspace{1.8cm}
+{1-\xi-\eta\over2\xi}\ln\eta\Big]
+\varepsilon\Phi_{_2}(\xi,\eta)+\cdots\Big\}\;,
\label{1LoopB0-49b}
\end{eqnarray}
where the first two terms are consistent with the well-known expression
of $B_{_0}$ function exactly~\cite{A.Denner93}.
Similarly the function $\Phi_{_2}$ in the kinematic region
$\lambda_{_{\xi,\eta}}^2\ge0$ is written as
\begin{eqnarray}
&&\Phi_{_2}(\xi,\eta)={\lambda_{_{\xi,\eta}}\over\xi}\ln\eta
\ln{\lambda_{_{\xi,\eta}}(1-\xi-\eta-\lambda_{_{\xi,\eta}})\over2\xi\eta}
-{1-\xi-\eta-\lambda_{_{\xi,\eta}}\over4\xi}\ln^2\eta
\nonumber\\
&&\hspace{2.0cm}
-{\lambda_{_{\xi,\eta}}\over\xi}\ln{1-\xi-\eta-\lambda_{_{\xi,\eta}}\over2\lambda_{_{\xi,\eta}}}
\ln{\lambda_{_{\xi,\eta}}(1-\xi-\eta-\lambda_{_{\xi,\eta}})\over2\xi\eta}
\nonumber\\
&&\hspace{2.0cm}
+{\lambda_{_{\xi,\eta}}\over\xi}\ln{1+\xi-\eta-\lambda_{_{\xi,\eta}}\over2\lambda_{_{\xi,\eta}}}
\ln{\lambda_{_{\xi,\eta}}(1-\xi-\eta-\lambda_{_{\xi,\eta}})\over\xi(1-\xi+\eta-\lambda_{_{\xi,\eta}})}
\nonumber\\
&&\hspace{2.0cm}
-{\lambda_{_{\xi,\eta}}\over\xi}L_{_{i_2}}({\lambda_{_{\xi,\eta}}(1-\xi-\eta-\lambda_{_{\xi,\eta}})\over2\xi\eta})
+{\lambda_{_{\xi,\eta}}\over\xi}L_{_{i_2}}({\lambda_{_{\xi,\eta}}(1-\xi-\eta-\lambda_{_{\xi,\eta}})
\over\xi(1-\xi+\eta-\lambda_{_{\xi,\eta}})})\;.
\label{1LoopB0-49c}
\end{eqnarray}
With the quadratic transformations in Eq.(\ref{1LoopB0-49}),
the $B_{_0}$ function is written as
\begin{eqnarray}
&&B_{_0}(p^2)=
{i\over(4\pi)^2}\Big({4\pi\mu^2\over m_{_2}^2}\Big)^\varepsilon
{\Gamma(1+\varepsilon)\over\varepsilon(1-2\varepsilon)}
\nonumber\\
&&\hspace{1.5cm}\times
\Big\{-{(1-\xi-\eta)\eta^{-\varepsilon}\over2\xi}
\;_{_2}F_{_1}\left(\left.\begin{array}{c}\varepsilon,1\\ {1\over2}+\varepsilon\end{array}\right|
-{\lambda_{_{\xi,\eta}}^2\over4\xi\eta}\right)\Big]
\nonumber\\
&&\hspace{1.5cm}
+{1+\xi-\eta\over2\xi}\;_{_2}F_{_1}\left(\left.\begin{array}{c}
\varepsilon,1\\ {1\over2}+\varepsilon\end{array}\right|
-{\lambda_{_{\xi,\eta}}^2\over4\xi}\right)\Big]\Big\}\;.
\label{1LoopB0-53}
\end{eqnarray}
in the kinematic region $m_{_2}^2>\max(m_{_1}^2,\;|p^2|)$, $\lambda_{_{\xi,\eta}}^2\le0$.
Using the expansion of Eq.(\ref{1LoopB0-51b}), one finally gets
\begin{eqnarray}
&&B_{_0}(p^2)\simeq
{i\Gamma(1+\varepsilon)\over(4\pi)^2(1-2\varepsilon)}
\Big({4\pi\mu^2\over m_{_2}^2}\Big)^\varepsilon
\nonumber\\
&&\hspace{1.5cm}\times
\Big\{{1\over\varepsilon}
+\Big[-{\lambda_{_{\xi,\eta}}\over\xi}\ln{\lambda_{_{\xi,\eta}}(1-\xi+\eta-\lambda_{_{\xi,\eta}})\over2\sqrt{\eta}}
+{1-\xi-\eta\over2\xi}\ln\eta\Big]
\nonumber\\
&&\hspace{1.5cm}
+\varepsilon\Phi_{_2}(\xi,\eta)+\cdots\Big\}\;,
\label{1LoopB0-53b}
\end{eqnarray}
where the function $\Phi_{_2}$ in the kinematic region
$\lambda_{_{\xi,\eta}}^2\le0$ is written as
\begin{eqnarray}
&&\Phi_{_2}(\xi,\eta)=-{1-\xi-\eta\over4\xi}\ln^2\eta
-{\sqrt{-\lambda_{_{\xi,\eta}}^2}\over\xi}\Big(\ln({-\lambda_{_{\xi,\eta}}^2\over\xi})
\arcsin({\xi+\eta-1\over2\sqrt{\xi\eta}})
\nonumber\\
&&\hspace{2.0cm}
+\ln({-\lambda_{_{\xi,\eta}}^2\over\xi})\arcsin({1+\xi-\eta\over2\sqrt{\xi}})
+{\rm Cl}_{_2}(2\arcsin({\xi+\eta-1\over2\sqrt{\xi\eta}}))\Big)
\nonumber\\
&&\hspace{2.0cm}
+{\rm Cl}_{_2}(2\arcsin({1+\xi-\eta\over2\sqrt{\xi}}))\Big)\;.
\label{1LoopB0-53c}
\end{eqnarray}

\section{The system of PDEs for two-loop vacuum\label{sec4}}
\indent\indent
Similarly the two-loop vacuum integral is written as the radial integral
of Bessel functions:
\begin{eqnarray}
&&V_{_2}=
{2^3(m_{_1}^2m_{_2}^2m_{_3}^2)^{D/2-1}\over(4\pi)^{D}(\mu^2)^{D-4}\Gamma(D/2)}
\int_0^\infty dx\Big({x\over2}\Big)^{D-1}k_{_{D/2-1}}(m_{_1}x)
k_{_{D/2-1}}(m_{_2}x)k_{_{D/2-1}}(m_{_3}x)\;.
\label{2LoopVacuum1-1}
\end{eqnarray}

Assuming $m_{_3}\ge\max(m_{_1},m_{_2})$,
we insert the power series
of $k_{_{D/2-1}}(m_{_1}x)$ and $k_{_{D/2-1}}(m_{_2}x)$ into Eq.(\ref{2LoopVacuum1-1}):
\begin{eqnarray}
&&V_{_2}=
{2(m_{_1}^2m_{_2}^2m_{_3}^2)^{D/2-1}\over(4\pi)^{D}(\mu^2)^{D-4}\Gamma(D/2)}
\Gamma^2(D/2-1)\Gamma^2(2-D/2)
\nonumber\\
&&\hspace{1.0cm}\times
\sum\limits_{n_{_1}=0}^\infty\sum\limits_{n_{_2}=0}^\infty
{1\over n_{_1}!n_{_2}!}
\int_0^\infty dx\Big({x\over2}\Big)^{D-1}k_{_{D/2-1}}(m_{_3}x)
\nonumber\\
&&\hspace{1.0cm}\times
\Big[-{1\over\Gamma(D/2+n_{_1})}\Big({m_{_1}x\over2}\Big)^{2n_{_1}}
+{1\over\Gamma(2-D/2+n_{_1})}\Big({m_{_1}x\over2}\Big)^{2(n_{_1}-D/2+1)}\Big]
\nonumber\\
&&\hspace{1.0cm}\times
\Big[-{1\over\Gamma(D/2+n_{_2})}\Big({m_{_2}x\over2}\Big)^{2n_{_2}}
+{1\over\Gamma(2-D/2+n_{_2})}\Big({m_{_2}x\over2}\Big)^{2(n_{_2}-D/2+1)}\Big]
\label{2LoopVacuum1-2}
\end{eqnarray}
Through the integral formulae in Eq.(\ref{base15}), the scalar integral is written as
\begin{eqnarray}
&&V_{_2}={1\over(4\pi)^{D}}\Big({m_{_3}^2\over\mu^2}\Big)^{D-3}
{\Gamma^2(3-{D\over2})\over(D-2)(D-3)}\varphi(s,t)
\label{2LoopVacuum1-3}
\end{eqnarray}
with $s={m_{_1}^2\over m_{_3}^2},\;t={m_{_2}^2\over m_{_3}^2}$.
Additionally, the function $\varphi(s,t)$ is defined as
\begin{eqnarray}
&&\varphi(s,t)=
-{2(D-3)\over(2-{D\over2})^2(1-{D\over2})}(st)^{D/2-1}F_{_4}
\left(\left.\begin{array}{cc}1,&{D\over2}\\
{D\over2},&{D\over2}\end{array}\right|s,\;t\right)
\nonumber\\
&&\hspace{1.7cm}
-{2\Gamma({D\over2}-1)\Gamma(4-D)\over\Gamma(3-{D\over2})}F_{_4}
\left(\left.\begin{array}{cc}3-D,&2-{D\over2}\\
2-{D\over2},&2-{D\over2}\end{array}\right|s,\;t\right)
\nonumber\\
&&\hspace{1.7cm}
+{2(D-3)\over(2-{D\over2})^2(1-{D\over2})}s^{D/2-1}F_{_4}
\left(\left.\begin{array}{cc}1,&2-{D\over2}\\
{D\over2},&2-{D\over2}\end{array}\right|s,\;t\right)
\nonumber\\
&&\hspace{1.7cm}
+{2(D-3)\over(2-{D\over2})^2(1-{D\over2})}t^{D/2-1}F_{_4}
\left(\left.\begin{array}{cc}1,&2-{D\over2}\\
2-{D\over2},&{D\over2}\end{array}\right|s,\;t\right)\;.
\label{2LoopVacuum1-4}
\end{eqnarray}
The expression of Eq.(\ref{2LoopVacuum1-3}) coincides with Eq.(4.3)
of Ref.~\cite{A.I.Davydychev93} from the MB method exactly.
Correspondingly, the double hypergeometric function
$\varphi(s,t)$ satisfies the system of homogeneous linear PDEs
\begin{eqnarray}
&&\Big\{(\hat{\vartheta}_{s}+\hat{\vartheta}_{t}+2-{D\over2})
(\hat{\vartheta}_{s}+\hat{\vartheta}_{t}+3-D)
-{1\over s}\hat{\vartheta}_{s}(\hat{\vartheta}_{s}+1
-{D\over2})\Big\}\varphi=0
\;,\nonumber\\
&&\Big\{(\hat{\vartheta}_{s}+\hat{\vartheta}_{t}+2-{D\over2})
(\hat{\vartheta}_{s}+\hat{\vartheta}_{t}+3-D)
-{1\over t}\hat{\vartheta}_{t}(\hat{\vartheta}_{t}+1
-{D\over2})\Big\}\varphi=0\;.
\label{2LoopVacuum1-5}
\end{eqnarray}
For the case $m_{_1}\ge\max(m_{_2},m_{_3})$, one similarly derives
\begin{eqnarray}
&&V_{_2}={1\over(4\pi)^{D}}\Big({m_{_1}^2\over\mu^2}\Big)^{D-3}
{\Gamma^2(3-{D\over2})\over(D-2)(D-3)}\varphi(s^\prime,t^\prime)
\label{2LoopVacuum1-6}
\end{eqnarray}
with $s^\prime={m_{_2}^2\over m_{_1}^2}={t\over s},\;
t^\prime={m_{_3}^2\over m_{_1}^2}={1\over s}$. We specify here
$\varphi(s^\prime,t^\prime)=s^{3-D}\varphi(s,t)$, which is
derived from the transformation of Apell functions~\cite{L.J.Slater66}
\begin{eqnarray}
&&F_{_4}\left(\left.\begin{array}{cc}a,&b\\
c_{_1},&c_{_2}\end{array}\right|s,\;t\right)
={\Gamma(c_{_2})\Gamma(b-a)\over\Gamma(b)\Gamma(c_{_2}-a)}(-t)^{-a}
F_{_4}\left(\left.\begin{array}{cc}a,&1+a-c_{_2}\\
c_{_1},&1+a-b\end{array}\right|{s\over t},\;{1\over t}\right)
\nonumber\\
&&\hspace{3.7cm}
+{\Gamma(c_{_2})\Gamma(a-b)\over\Gamma(a)\Gamma(c_{_2}-b)}(-t)^{-b}
F_{_4}\left(\left.\begin{array}{cc}b,&1+b-c_{_2}\\
c_{_1},&1-a+b\end{array}\right|{s\over t},\;{1\over t}\right)\;.
\label{2LoopVacuum1-6-1}
\end{eqnarray}
Using the reduction formulae above, we get the well-known results of Ref.~\cite{A.I.Davydychev93}
\begin{eqnarray}
&&V_{_2}={\Gamma^2(1+\varepsilon)\over2(4\pi)^4(1-\varepsilon)(1-2\varepsilon)}
\Big({4\pi\mu^2\over m_{_3}^2}\Big)^{2\varepsilon}m_{_3}^2\Big\{
-{1\over\varepsilon^2}(1+s+t)+{2\over\varepsilon}(s\ln s+t\ln t)
\nonumber\\
&&\hspace{0.9cm}
-s\ln^2s-t\ln^2t+(1-s-t)\ln s\ln t-\lambda_{_{s,t}}\Phi(s,t)\Big\}\;,
\label{2LoopVacuum1-7}
\end{eqnarray}
where $\lambda_{_{s,t}}=1+s^2+t^2-2s-2t-2st$, and the concrete expression
of $\Phi(s,t)$ can be found in Ref.~\cite{A.I.Davydychev93}.

\section{The system of PDEs for scalar integral
from two-loop sunset diagram\label{sec5}}
\indent\indent
In order to obtain the multiple hypergeometric functions of
certain connected regions of independent kinematic variables, we present
the scalar integral of two-loop sunset diagram as
the radial integral of the modified Bessel functions:
\begin{eqnarray}
&&\Sigma_{_{\ominus}}(p^2)={8(m_{_1}^2m_{_2}^2m_{_3}^2)^{D/2-1}\over(4\pi)^D}\mu^{2(4-D)}
\int_0^\infty dx\Big({x\over2}\Big)^{D-1}j_{_{D/2-1}}(p_{_{\rm E}}x)k_{_{D/2-1}}(m_{_1}x)
\nonumber\\
&&\hspace{1.9cm}\times
k_{_{D/2-1}}(m_{_2}x)k_{_{D/2-1}}(m_{_3}x)\;.
\label{equivalence17-a}
\end{eqnarray}

Inserting the power series of
$k_{_{D/2-1}}(m_{_i}x)\;(i=1,\;2,\;3)$ into Eq.(\ref{equivalence17-a}),
one obtains
\begin{eqnarray}
&&\Sigma_{_{\ominus}}(p^2)=
{p_{_{\rm E}}^2\over(4\pi)^4}\Big({4\pi\mu^2\over p_{_{\rm E}}^2}\Big)^{2\varepsilon}
\Big({m_{_1}^2m_{_2}^2m_{_3}^2\over p_{_{\rm E}}^6}\Big)^{1-\varepsilon}
\Big[\Gamma^2(1-\varepsilon)\Gamma^2(\varepsilon)\Big]
\sum\limits_{n_{_1}=0}^\infty\sum\limits_{n_{_2}=0}^\infty\sum\limits_{n_{_3}=0}^\infty
\nonumber\\
&&\hspace{1.9cm}\times
\Big\{{(-)^{n_{_1}+n_{_2}+n_{_3}}\Gamma(1+n_{_1}+n_{_2}+n_{_3})
\Gamma(\varepsilon+n_{_1}+n_{_2}+n_{_3})\over n_{_1}!n_{_2}!n_{_3}!
\Gamma(2-\varepsilon+n_{_1})\Gamma(\varepsilon+n_{_2})\Gamma(2-\varepsilon+n_{_3})}
\nonumber\\
&&\hspace{1.9cm}\times
\Big({m_{_1}^2\over p_{_{\rm E}}^2}\Big)^{n_{_1}}
\Big({m_{_2}^2\over p_{_{\rm E}}^2}\Big)^{n_{_2}-1+\varepsilon}
\Big({m_{_3}^2\over p_{_{\rm E}}^2}\Big)^{n_{_3}}
\nonumber\\
&&\hspace{1.9cm}
+{(-)^{n_{_1}+n_{_2}+n_{_3}}\Gamma(1+n_{_1}+n_{_2}+n_{_3})
\Gamma(\varepsilon+n_{_1}+n_{_2}+n_{_3})\over n_{_1}!n_{_2}!n_{_3}!
\Gamma(\varepsilon+n_{_1})\Gamma(2-\varepsilon+n_{_2})\Gamma(2-\varepsilon+n_{_3})}
\nonumber\\
&&\hspace{1.9cm}\times
\Big({m_{_1}^2\over p_{_{\rm E}}^2}\Big)^{n_{_1}-1+\varepsilon}
\Big({m_{_2}^2\over p_{_{\rm E}}^2}\Big)^{n_{_2}}
\Big({m_{_3}^2\over p_{_{\rm E}}^2}\Big)^{n_{_3}}
\nonumber\\
&&\hspace{1.9cm}
+{(-)^{n_{_1}+n_{_2}+n_{_3}}\Gamma(1+n_{_1}+n_{_2}+n_{_3})
\Gamma(\varepsilon+n_{_1}+n_{_2}+n_{_3})\over n_{_1}!n_{_2}!n_{_3}!
\Gamma(2-\varepsilon+n_{_1})\Gamma(2-\varepsilon+n_{_2})\Gamma(\varepsilon+n_{_3})}
\nonumber\\
&&\hspace{1.9cm}\times
\Big({m_{_1}^2\over p_{_{\rm E}}^2}\Big)^{n_{_1}}\Big({m_{_2}^2\over p_{_{\rm E}}^2}\Big)^{n_{_2}}
\Big({m_{_3}^2\over p_{_{\rm E}}^2}\Big)^{n_{_3}-1+\varepsilon}
\nonumber\\
&&\hspace{1.9cm}
+{(-)^{n_{_1}+n_{_2}+n_{_3}}\Gamma(\varepsilon+n_{_1}+n_{_2}+n_{_3})
\Gamma(-1+2\varepsilon+n_{_1}+n_{_2}+n_{_3})\over n_{_1}!n_{_2}!n_{_3}!
\Gamma(\varepsilon+n_{_1})\Gamma(\varepsilon+n_{_2})\Gamma(2-\varepsilon+n_{_3})}
\nonumber\\
&&\hspace{1.9cm}\times
{\Gamma(\varepsilon)\Gamma(1-\varepsilon)\over\Gamma(2\varepsilon)\Gamma(1-2\varepsilon)}
\Big({m_{_1}^2\over p_{_{\rm E}}^2}\Big)^{n_{_1}-1+\varepsilon}
\Big({m_{_2}^2\over p_{_{\rm E}}^2}\Big)^{n_{_2}-1+\varepsilon}
\Big({m_{_3}^2\over p_{_{\rm E}}^2}\Big)^{n_{_3}}
\nonumber\\
&&\hspace{1.9cm}
+{(-)^{n_{_1}+n_{_2}+n_{_3}}\Gamma(\varepsilon+n_{_1}+n_{_2}+n_{_3})
\Gamma(-1+2\varepsilon+n_{_1}+n_{_2}+n_{_3})\over n_{_1}!n_{_2}!n_{_3}!
\Gamma(2-\varepsilon+n_{_1})\Gamma(\varepsilon+n_{_2})\Gamma(\varepsilon+n_{_3})}
\nonumber\\
&&\hspace{1.9cm}\times
{\Gamma(\varepsilon)\Gamma(1-\varepsilon)\over\Gamma(2\varepsilon)\Gamma(1-2\varepsilon)}
\Big({m_{_1}^2\over p_{_{\rm E}}^2}\Big)^{n_{_1}}\Big({m_{_2}^2\over p_{_{\rm E}}^2}\Big)^{n_{_2}-1+\varepsilon}
\Big({m_{_3}^2\over p_{_{\rm E}}^2}\Big)^{n_{_3}-1+\varepsilon}
\nonumber\\
&&\hspace{1.9cm}
+{(-)^{n_{_1}+n_{_2}+n_{_3}}\Gamma(\varepsilon+n_{_1}+n_{_2}+n_{_3})
\Gamma(-1+2\varepsilon+n_{_1}+n_{_2}+n_{_3})\over n_{_1}!n_{_2}!n_{_3}!
\Gamma(\varepsilon+n_{_1})\Gamma(2-\varepsilon+n_{_2})\Gamma(\varepsilon+n_{_3})}
\nonumber\\
&&\hspace{1.9cm}\times
{\Gamma(\varepsilon)\Gamma(1-\varepsilon)\over\Gamma(2\varepsilon)\Gamma(1-2\varepsilon)}
\Big({m_{_1}^2\over p_{_{\rm E}}^2}\Big)^{n_{_1}-1+\varepsilon}
\Big({m_{_2}^2\over p_{_{\rm E}}^2}\Big)^{n_{_2}}
\Big({m_{_3}^2\over p_{_{\rm E}}^2}\Big)^{n_{_3}-1+\varepsilon}
\nonumber\\
&&\hspace{1.9cm}
+{(-)^{n_{_1}+n_{_2}+n_{_3}}\Gamma(-1+2\varepsilon+n_{_1}+n_{_2}+n_{_3})
\Gamma(-2+3\varepsilon+n_{_1}+n_{_2}+n_{_3})\over n_{_1}!n_{_2}!n_{_3}!
\Gamma(\varepsilon+n_{_1})\Gamma(\varepsilon+n_{_2})\Gamma(\varepsilon+n_{_3})}
\nonumber\\
&&\hspace{1.9cm}\times
{\Gamma(\varepsilon)\Gamma(1-\varepsilon)\over\Gamma(3\varepsilon)\Gamma(1-3\varepsilon)}
\Big({m_{_1}^2\over p_{_{\rm E}}^2}\Big)^{n_{_1}-1+\varepsilon}
\Big({m_{_2}^2\over p_{_{\rm E}}^2}\Big)^{n_{_2}-1+\varepsilon}
\Big({m_{_3}^2\over p_{_{\rm E}}^2}\Big)^{n_{_3}-1+\varepsilon}\Big\}\;,
\label{sunset2}
\end{eqnarray}
where $p_{_{\rm E}}^2$ represents the momentum squared in Euclidean space.
Substituting $p_{_{\rm E}}^2\rightarrow-p^2$, we get the scalar integral as
\begin{eqnarray}
&&\Sigma_{_{\ominus}}(p^2)=
-{p^2\over(4\pi)^4}\Big({4\pi\mu^2\over-p^2}\Big)^{2\varepsilon}
\nonumber\\
&&\hspace{1.9cm}\times
\Big\{{\Gamma^2(\varepsilon)\over(1-\varepsilon)^2}
(x_{_1}x_{_2})^{1-\varepsilon}
F_{_C}^{(3)}\left(\left.\begin{array}{c}1,\varepsilon\;\\
2-\varepsilon,2-\varepsilon,\varepsilon\end{array}\right|
x_{_1},x_{_2},x_{_3}\right)
\nonumber\\
&&\hspace{1.9cm}
+{\Gamma^2(\varepsilon)\over(1-\varepsilon)^2}
(x_{_2}x_{_3})^{1-\varepsilon}
F_{_C}^{(3)}\left(\left.\begin{array}{c}1,\varepsilon\;\\
\varepsilon,2-\varepsilon,2-\varepsilon\end{array}\right|
x_{_1},x_{_2},x_{_3}\right)
\nonumber\\
&&\hspace{1.9cm}
+{\Gamma^2(\varepsilon)\over(1-\varepsilon)^2}
(x_{_1}x_{_3})^{1-\varepsilon}
F_{_C}^{(3)}\left(\left.\begin{array}{c}1,\varepsilon\;\\
2-\varepsilon,\varepsilon,2-\varepsilon\end{array}\right|
x_{_1},x_{_2},x_{_3}\right)
\nonumber\\
&&\hspace{1.9cm}
-{\Gamma^2(1-\varepsilon)\Gamma^2(\varepsilon)\over(1-\varepsilon)\Gamma(2-2\varepsilon)}
(-x_{_1})^{1-\varepsilon}
F_{_C}^{(3)}\left(\left.\begin{array}{c}2\varepsilon-1,\varepsilon\;\\
2-\varepsilon,\varepsilon,\varepsilon\end{array}\right|
x_{_1},x_{_2},x_{_3}\right)
\nonumber\\
&&\hspace{1.9cm}
-{\Gamma^2(1-\varepsilon)\Gamma^2(\varepsilon)\over(1-\varepsilon)\Gamma(2-2\varepsilon)}
(-x_{_2})^{1-\varepsilon}
F_{_C}^{(3)}\left(\left.\begin{array}{c}2\varepsilon-1,\varepsilon\;\\
\varepsilon,2-\varepsilon,\varepsilon\end{array}\right|
x_{_1},x_{_2},x_{_3}\right)
\nonumber\\
&&\hspace{1.9cm}
-{\Gamma^2(1-\varepsilon)\Gamma^2(\varepsilon)\over(1-\varepsilon)\Gamma(2-2\varepsilon)}
(-x_{_3})^{1-\varepsilon}
F_{_C}^{(3)}\left(\left.\begin{array}{c}2\varepsilon-1,\varepsilon\;\\
\varepsilon,\varepsilon,2-\varepsilon\end{array}\right|
x_{_1},x_{_2},x_{_3}\right)
\nonumber\\
&&\hspace{1.9cm}
+{\Gamma^3(1-\varepsilon)\Gamma(-1+2\varepsilon)\over\Gamma(3-3\varepsilon)}
F_{_C}^{(3)}\left(\left.\begin{array}{c}2\varepsilon-1,3\varepsilon-2\;\\
\varepsilon,\varepsilon,\varepsilon\end{array}\right|
x_{_1},x_{_2},x_{_3}\right)\Big\}
\nonumber\\
&&\hspace{1.5cm}=
-{p^2\over(4\pi)^4}\Big({4\pi\mu^2\over-p^2}\Big)^{4-D}
\Gamma^2(3-{D\over2})T_{_{123}}^{p}(x_{_1},x_{_2},x_{_3})\;.
\label{sunset3}
\end{eqnarray}
Here $x_{_1}=m_{_1}^2/p^2,\;x_{_2}=m_{_2}^2/p^2,\;x_{_3}=m_{_3}^2/p^2$,
$F_{_C}^{(3)}$ is the Lauricella function of three independent variables
\begin{eqnarray}
&&F_{_C}^{(3)}\left(\left.\begin{array}{c}a,b\;\\
c_{_1},c_{_2},c_{_3}\end{array}\right|x,\;y,\;z\right)
=\sum\limits_{n_{_x}=0}^\infty\sum\limits_{n_{_y}=0}^\infty
\sum\limits_{n_{_z}=0}^\infty{(a)_{_{n_{_x}+n_{_y}+n_{_z}}}(b)_{_{n_{_x}+n_{_y}+n_{_z}}}
\over n_{_x}!n_{_y}!n_{_z}!(c_{_1})_{_{n_{_x}}}(c_{_2})_{_{n_{_y}}}(c_{_3})_{_{n_{_z}}}}
x^{n_{_x}}y^{n_{_y}}z^{n_{_z}}
\label{Lauricella}
\end{eqnarray}
which is convergent in the connected region $\sqrt{|x_{_1}|}+\sqrt{|x_{_2}|}+\sqrt{|x_{_3}|}\le1$.
Obviously the function $T_{_{123}}^{p}$ satisfies the system of homogeneous linear PDEs
\begin{eqnarray}
&&\Big\{(\sum\limits_{i=1}^3\hat{\vartheta}_{x_{_i}}+3-D)
(\sum\limits_{i=1}^3\hat{\vartheta}_{x_{_i}}+4-{3D\over2})
-{1\over x_{_1}}\hat{\vartheta}_{x_{_1}}(\hat{\vartheta}_{x_{_1}}+1
-{D\over2})\Big\}T_{_{123}}^{p}=0
\;,\nonumber\\
&&\Big\{(\sum\limits_{i=1}^3\hat{\vartheta}_{x_{_i}}+3-D)
(\sum\limits_{i=1}^3\hat{\vartheta}_{x_{_i}}+4-{3D\over2})
-{1\over x_{_2}}\hat{\vartheta}_{x_{_2}}(\hat{\vartheta}_{x_{_2}}+1
-{D\over2})\Big\}T_{_{123}}^{p}=0
\;,\nonumber\\
&&\Big\{(\sum\limits_{i=1}^3\hat{\vartheta}_{x_{_i}}+3-D)
(\sum\limits_{i=1}^3\hat{\vartheta}_{x_{_i}}+4-{3D\over2})
-{1\over x_{_3}}\hat{\vartheta}_{x_{_3}}(\hat{\vartheta}_{x_{_3}}+1
-{D\over2})\Big\}T_{_{123}}^{p}=0\;.
\label{sunset5}
\end{eqnarray}

Similarly inserting the power series of $j_{_{D/2-1}}(p_{_{\rm E}}x)$,
$k_{_{D/2-1}}(m_{_1}x)$, $k_{_{D/2-1}}(m_{_2}x)$ into Eq.(\ref{equivalence17-a})
in the kinematic region $m_{_3}^2>\max(|p^2|,m_{_1}^2,m_{_2}^2)$, one obtains
\begin{eqnarray}
&&\Sigma_{_{\ominus}}(p^2)=
{m_{_3}^2\over(4\pi)^4}\Big({4\pi\mu^2\over m_{_3}^2}\Big)^{2\varepsilon}
\nonumber\\
&&\hspace{1.9cm}\times
\Big\{{\Gamma^2(\varepsilon)\over(1-\varepsilon)^2}
\Big({m_{_1}^2m_{_2}^2\over m_{_3}^2}\Big)^{1-\varepsilon}
F_{_C}^{(3)}\left(\left.\begin{array}{c}1,2-\varepsilon\;\\
2-\varepsilon,2-\varepsilon,2-\varepsilon\end{array}\right|
\xi_{_1},\;\xi_{_2},\;\xi_{_3}\right)
\nonumber\\
&&\hspace{1.9cm}
-{\Gamma^2(\varepsilon)\over(1-\varepsilon)^2}
\Big({m_{_1}^2\over m_{_3}^2}\Big)^{1-\varepsilon}
F_{_C}^{(3)}\left(\left.\begin{array}{c}1,\varepsilon\;\\
2-\varepsilon,\varepsilon,2-\varepsilon\end{array}\right|
\xi_{_1},\;\xi_{_2},\;\xi_{_3}\right)
\nonumber\\
&&\hspace{1.9cm}
-{\Gamma^2(\varepsilon)\over(1-\varepsilon)^2}
\Big({m_{_2}^2\over m_{_3}^2}\Big)^{1-\varepsilon}
F_{_C}^{(3)}\left(\left.\begin{array}{c}1,\varepsilon\;\\
\varepsilon,2-\varepsilon,2-\varepsilon\end{array}\right|
\xi_{_1},\;\xi_{_2},\;\xi_{_3}\right)
\nonumber\\
&&\hspace{1.9cm}
+{\Gamma(\varepsilon)\Gamma(2\varepsilon-1)\Gamma(1-\varepsilon)\over1-\varepsilon}
F_{_C}^{(3)}\left(\left.\begin{array}{c}\varepsilon,2\varepsilon-1\;\\
\varepsilon,\varepsilon,2-\varepsilon\end{array}\right|
\xi_{_1},\;\xi_{_2},\;\xi_{_3}\right)\Big\}
\nonumber\\
&&\hspace{1.5cm}=
{m_{_3}^2\over(4\pi)^4}\Big({4\pi\mu^2\over m_{_3}^2}\Big)^{4-D}
\Gamma^2(3-{D\over2})T_{_{123}}^{m}(\xi_{_1},\;\xi_{_2},\;\xi_{_3})\;,
\label{sunset8}
\end{eqnarray}
with $\xi_{_1}=m_{_1}^2/m_{_3}^2,\;\xi_{_2}=m_{_2}^2/m_{_3}^2,\;\xi_{_3}=p^2/m_{_3}^2$,
and the convergent region of the triple hypergeometric function is
$\sqrt{|\xi_{_1}|}+\sqrt{|\xi_{_2}|}+\sqrt{|\xi_{_3}|}\le1$, i.e.
$1+\sqrt{|x_{_1}|}+\sqrt{|x_{_2}|}\le\sqrt{|x_{_3}|}$.
We specify here that the expression of Eq.(\ref{sunset8}) can be obtained
equivalently through the MB method~\cite{F.A.Berends94}. In fact we recover
the triple  hypergeometric functions of Eq.(\ref{sunset3}) from Eq.(\ref{sunset8})
through the transformation of Lauricella functions
\begin{eqnarray}
&&F_{_C}^{(3)}\left(\left.\begin{array}{c}a,b,\\
c_{_1},c_{_2},c_{_3}\end{array}\right|s,\;t,\;u\right)
\nonumber\\
&&\hspace{-0.5cm}
={\Gamma(c_{_3})\Gamma(b-a)\over\Gamma(b)\Gamma(c_{_3}-a)}(-t)^{-a}
F_{_C}^{(3)}\left(\left.\begin{array}{c}a,1+a-c_{_3},\\
c_{_1},c_{_2},1+a-b\end{array}\right|{s\over u},\;{t\over u},\;{1\over u}\right)
\nonumber\\
&&\hspace{-0.2cm}
+{\Gamma(c_{_3})\Gamma(a-b)\over\Gamma(a)\Gamma(c_{_3}-b)}(-t)^{-b}
F_{_C}^{(3)}\left(\left.\begin{array}{c}b,1+b-c_{_2},\\
c_{_1},c_{_2},1-a+b\end{array}\right|{s\over u},\;{t\over u},\;{1\over u}\right)\;.
\label{Lauricella1}
\end{eqnarray}
Additionally, the function $T_{_{123}}^{m}$ satisfies the system
of homogeneous linear PDEs
\begin{eqnarray}
&&\Big\{(\sum\limits_{i=1}^3\hat{\vartheta}_{\xi_{_i}}+3-D)
(\sum\limits_{i=1}^3\hat{\vartheta}_{\xi_{_i}}+2-{D\over2})
-{1\over\xi_{_1}}\hat{\vartheta}_{\xi_{_1}}(\hat{\vartheta}_{\xi_{_1}}+1
-{D\over2})\Big\}T_{_{123}}^{m}=0
\;,\nonumber\\
&&\Big\{(\sum\limits_{i=1}^3\hat{\vartheta}_{\xi_{_i}}+3-D)
(\sum\limits_{i=1}^3\hat{\vartheta}_{\xi_{_i}}+2-{D\over2})
-{1\over\xi_{_2}}\hat{\vartheta}_{\xi_{_2}}(\hat{\vartheta}_{\xi_{_2}}+1
-{D\over2})\Big\}T_{_{123}}^{m}=0
\;,\nonumber\\
&&\Big\{(\sum\limits_{i=1}^3\hat{\vartheta}_{\xi_{_i}}+3-D)
(\sum\limits_{i=1}^3\hat{\vartheta}_{\xi_{_i}}+2-{D\over2})
-{1\over\xi_{_3}}\hat{\vartheta}_{\xi_{_3}}(\hat{\vartheta}_{\xi_{_3}}-1
+{D\over2})\Big\}T_{_{123}}^{m}=0\;.
\label{sunset10}
\end{eqnarray}
Interchanging $m_{_3}\leftrightarrow m_{_1}$ and $m_{_3}\leftrightarrow m_{_2}$
in the triple hypergeometric functions of Eq.(\ref{sunset8}) and the system of PDEs
of Eq.(\ref{sunset10}), one obtains the corresponding results of the kinematic regions
$m_{_1}^2>\max(|p^2|,m_{_2}^2,\;m_{_3}^2)$ and $m_{_2}^2>\max(|p^2|,m_{_1}^2,\;m_{_3}^2)$, respectively.
A point specified here is that the system of homogeneous linear PDEs of
Eq.(\ref{sunset5}) is equivalent to that of Eq.(\ref{sunset10}).
Inserting $T_{_{123}}^{m}(\xi_{_1},\;\xi_{_2},\;\xi_{_3})=(-x_{_3})^{3-D}T_{_{123}}^{p}(x_{_1},x_{_2},x_{_3})$,
$\xi_{_1}=x_{_1}/x_{_3},\;\xi_{_2}=x_{_2}/x_{_3},\;\xi_{_3}=1/x_{_3}$ into Eq.(\ref{sunset10}),
one derives three linear combinations of PDEs in Eq.(\ref{sunset5})
explicitly. This implicates that the function defined through Eq.(\ref{sunset3})
satisfies the system of PDEs in Eq.(\ref{sunset5}). In other words, the continuation of
$T_{_{123}}^{p}(x_{_1},x_{_2},x_{_3})$ from its convergent regions
to the whole kinematic domain can be made numerically through the system of homogeneous linear PDEs.
We will address this issue in detail in section \ref{sec7}.

\section{The systems of PDEs for one-loop 3-point diagram\label{sec6}}
\indent\indent
The hypergeometric function method can also be applied to analyze the scalar
integrals for one-loop 3-point or 4-point diagrams. For simplification,
we present the result of the massless one-loop 3-point diagram here:
\begin{eqnarray}
&&C_{_0}=\int{d^Dq\over(2\pi)^D}{1\over q^2(q+p_{_1})^2
(q-p_{_2})^2}
\nonumber\\
&&\hspace{0.5cm}=
-i{2^{3(D-2)}\Gamma^3(D/2-1)\over(4\pi)^{3D/2}}
\int d^Dx_{_1}d^Dx_{_2}{\exp\{i({\bf x}_{_1}\cdot{\bf p}_{_{\rm 1E}}+{\bf x}_{_2}\cdot{\bf p}_{_{\rm 2E}})\}
\over x_{_1}^{D-2}x_{_2}^{D-2}|{\bf x}_{_1}-{\bf x}_{_2}|^{D-2}}\;.
\label{1LoopC01-a}
\end{eqnarray}
Using the generating function of Gegenbauer's polynomials
\begin{eqnarray}
&&{1\over|{\bf x}-{\bf x}^\prime|^{2\mu}}=\sum\limits_{n=0}^\infty
C_{_n}^\mu(\hat{\bf x}\cdot\hat{\bf x}^\prime)\Big[{x^n\over x^{\prime(n+2\mu)}}
\Theta(x^\prime-x)+{x^{\prime n}\over x^{(n+2\mu)}}\Theta(x-x^\prime)\Big]\;,
\label{base7}
\end{eqnarray}
and the orthogonality of Gegenbauer's polynomials~\cite{H.Bateman53},
one writes the massless one-loop 3-point function as
\begin{eqnarray}
&&C_{_0}=-i{2^{D-2}\Gamma^3(D/2-1)\over(4\pi)^{D/2}p_{_{\rm 1E}}p_{_{\rm 2E}}}
\sum\limits_{n=0}^{\infty}(-)^n
C_{_{n}}^{D/2-1}(\hat{p}_{_{\rm 1E}}\cdot\hat{p}_{_{\rm 2E}})
\nonumber\\
&&\hspace{1.0cm}\times
\int dx_{_1}dx_{_2}\Big({x_{_1}p_{_{1E}}\over2}\Big)^{n+1}
\Big({x_{_2}p_{_{2E}}\over2}\Big)^{n+1}
j_{_{D/2-1+n}}(x_{_1}p_{_{\rm 1E}})j_{_{D/2-1+n}}(x_{_2}p_{_{\rm 2E}})
\nonumber\\
&&\hspace{1.0cm}\times
\Big[{x_{_1}^{n}\over x_{_2}^{(n+D-2)}}
\Theta(x_{_2}-x_{_1})+{x_{_2}^{n}\over x_{_1}^{(n+D-2)}}\Theta(x_{_1}-x_{_2})\Big]\;,
\label{1LoopC01}
\end{eqnarray}
where $\Theta(t)$ denotes the step function, and $C_{_n}^\mu(t)$ is the
Gegenbauer's polynomial, respectively. In the kinematic region
$|p_{_2}^2|\ge\max(|p_{_1}^2|,\;|(p_{_1}-p_{_2})^2|)$,
the radial integral is transformed as
\begin{eqnarray}
&&\hspace{1.0cm}
\int dx_{_1}dx_{_2}\Big({x_{_1}p_{_{\rm 1E}}\over2}\Big)^{n+1}
\Big({x_{_2}p_{_{2E}}\over2}\Big)^{n+1}
j_{_{D/2-1+n}}(x_{_1}p_{_{\rm 1E}})j_{_{D/2-1+n}}(x_{_2}p_{_{2E}})
\nonumber\\
&&\hspace{1.0cm}\times
\Big[{x_{_1}^{n}\over x_{_2}^{(n+D-2)}}
\Theta(x_{_2}-x_{_1})+{x_{_2}^{n}\over x_{_1}^{(n+D-2)}}\Theta(x_{_1}-x_{_2})\Big]
\nonumber\\
&&\hspace{0.5cm}=
{1\over2^{D-2}}
\Big\{p_{_{\rm 1E}}^{-n-1}p_{_{\rm 2E}}^{n+D-3}
\int_0^\infty dt_{_2} \Big({t_{_2}\over2}\Big)^{3-D}
\nonumber\\
&&\hspace{1.0cm}\times
j_{_{D/2-1+n}}(t_{_2})
\int_0^{p_{_{\rm 1E}}t_{_2}/p_{_{\rm 2E}}} dt_{_1}\Big({t_{_1}\over2}\Big)^{2n+1}
j_{_{D/2-1+n}}(t_{_1})
\nonumber\\
&&\hspace{1.0cm}
+p_{_{\rm 1E}}^{n+D-3}p_{_{\rm 2E}}^{-n-1}
\int_0^\infty dt_{_2}\Big({t_{_2}\over2}\Big)^{2n+1}
j_{_{D/2-1+n}}(t_{_2})
\nonumber\\
&&\hspace{1.0cm}\times
\Big[\int_0^\infty dt_{_1}-\int_0^{p_{_{\rm 1E}}t_{_2}/p_{_{\rm2E}}} dt_{_1}\Big]
\Big({t_{_1}\over2}\Big)^{3-D}j_{_{D/2-1+n}}(t_{_1})\Big\}\;.
\label{1LoopC04}
\end{eqnarray}
With Eq.(\ref{base15}) the scalar integral is rewritten as
\begin{eqnarray}
&&C_{_0}=
-i{p_{_{\rm 2E}}^{D-6}\Gamma^3(D/2-1)\over(4\pi)^{D/2}}
\sum\limits_{n=0}^{\infty}(-)^n
C_{_{n}}^{D/2-1}(\hat{p}_{_{\rm 1E}}\cdot\hat{p}_{_{\rm 2E}})
\nonumber\\
&&\hspace{1.0cm}\times
\Big\{{\Gamma(2-{D\over2})\over\Gamma({D\over2}-1)}
{\Gamma(1+n)\over\Gamma(D-2+n)}\Big({p_{_{\rm1E}}\over p_{_{\rm2E}}}\Big)^{D-4+n}
\nonumber\\
&&\hspace{1.0cm}
+2\sum\limits_{q=0}^\infty
{(-)^{q}\Gamma(3-{D\over2}+q+n)\over q!\Gamma(q+n+{D\over2})\Gamma(D-3-q)}
\nonumber\\
&&\hspace{1.0cm}\times
\Big[{1\over2q+2n+2}-{1\over2q-D+4}\Big]
\Big({p_{_{\rm1E}}\over p_{_{\rm2E}}}\Big)^{2q+n}\Big\}\;.
\label{1LoopC05}
\end{eqnarray}
Taking the concrete expressions of Gegenbauer's polynomials
\begin{eqnarray}
&&C_{_{2n}}^\mu(t)={(-)^n(\mu)_n\over n!}
\;_{_2}F_{_1}\left(\left.\begin{array}{c}-n,\;\mu+n\\ {1\over2}\end{array}\right|
t^2\right)
\;,\nonumber\\
&&C_{_{2n+1}}^\mu(t)={(-)^n(\mu)_{n+1}\over n!}2t
\;_{_2}F_{_1}\left(\left.\begin{array}{c}-n,\;1+\mu+n\\ {3\over2}\end{array}\right|
t^2\right)
\label{1LoopC07}
\end{eqnarray}
and substituting $p_{_{\rm1E}}^2\rightarrow-p_{_1}^2$,
$p_{_{\rm2E}}^2\rightarrow-p_{_2}^2$, $p_{_{\rm 1E}}\cdot p_{_{\rm 2E}}\rightarrow
-p_{_1}\cdot p_{_2}$, we present the massless one-loop 3-point function as
\begin{eqnarray}
&&C_{_0}=-i{\Gamma({1\over2})\Gamma^3(D/2-1)\over(4\pi)^{D/2}(-p_{_2}^2)^{3-D/2}}
\Big\{C_{_0}^{(1)}({p_{_1}^2\over p_{_2}^2},{(p_{_1}\cdot p_{_2})^2\over p_{_1}^2p_{_2}^2})
\nonumber\\
&&\hspace{1.0cm}
+C_{_0}^{(2)}({p_{_1}^2\over p_{_2}^2},{p_{_1}^2\over p_{_2}^2},{(p_{_1}\cdot p_{_2})^2\over p_{_1}^2p_{_2}^2})
+C_{_0}^{(3)}({p_{_1}^2\over p_{_2}^2},{p_{_1}^2\over p_{_2}^2},
{(p_{_1}\cdot p_{_2})^2\over p_{_1}^2p_{_2}^2})\Big\}
\label{1LoopC09-a}
\end{eqnarray}
with
\begin{eqnarray}
&&C_{_0}^{(1)}(u,v)
=u^{D/2-2}{\Gamma(2-{D\over2})\over\Gamma({D\over2}-1)}
\sum\limits_{n=0}^{\infty}\sum\limits_{r=0}^{n}{(-)^n\Gamma(-n+r)\over n!r!\Gamma(-n)}
\nonumber\\
&&\hspace{2.2cm}\times
\Big\{{\Gamma({D\over2}-1+n+r)\Gamma(1+2n)\over\Gamma(D-2+2n)
\Gamma({1\over2}+r)}u^nv^r
\nonumber\\
&&\hspace{2.2cm}
-{\Gamma({D\over2}+n+r)\Gamma(2+2n)\over\Gamma(D-1+2n)
\Gamma({3\over2}+r)}u^{n+1/2}v^{r+1/2}\Big\}
\;,\nonumber\\
&&C_{_0}^{(2)}(u,u^\prime,v)=2u^{-2+D/2}u^{\prime\;2-D/2}
\sum\limits_{n=0}^{\infty}\sum\limits_{q=0}^{\infty}
\sum\limits_{r=0}^{n}
\nonumber\\
&&\hspace{2.8cm}\times
{(-)^{n+q}\Gamma(-n+r)\over n!q!r!\Gamma(D-3-q)\Gamma(-n)}
\nonumber\\
&&\hspace{2.8cm}\times
\Big\{{\Gamma(3-{D\over2}+q+2n)\Gamma({D\over2}-1+n+r)\over
(2q+4n+2)\Gamma({D\over2}+q+2n)\Gamma({1\over2}+r)}
u^nu^{\prime\;q}v^r
\nonumber\\
&&\hspace{2.8cm}
-{\Gamma(4-{D\over2}+q+2n)\Gamma({D\over2}+n+r)\over
(2q+4n+4)\Gamma(1+{D\over2}+q+2n)\Gamma({3\over2}+r)}
u^{n+1/2}u^{\prime\;q}v^{r+1/2}\Big\}
\;,\nonumber\\
&&C_{_0}^{(3)}(u,u^\prime,v)=-2u^{-2+D/2}u^{\prime\;2-D/2}
\sum\limits_{n=0}^{\infty}\sum\limits_{q=0}^{\infty}\sum\limits_{r=0}^{n}
\nonumber\\
&&\hspace{2.8cm}\times
{(-)^{n+q}\Gamma(-n+r)\over n!q!r!(2q-D+4)\Gamma(D-3-q)\Gamma(-n)}
\nonumber\\
&&\hspace{2.8cm}\times
\Big\{{\Gamma(3-{D\over2}+q+2n)\Gamma({D\over2}-1+n+r)\over
\Gamma({D\over2}+q+2n)\Gamma({1\over2}+r)}u^nu^{\prime\;q}v^r
\nonumber\\
&&\hspace{2.8cm}
-{\Gamma(4-{D\over2}+q+2n)\Gamma({D\over2}+n+r)\over
\Gamma(1+{D\over2}+q+2n)\Gamma({3\over2}+r)}u^{n+1/2}u^{\prime\;q}v^{r+1/2}\Big\}\;.
\label{1LoopC09}
\end{eqnarray}
The system of PDEs satisfied by the first term is written explicitly as
\begin{eqnarray}
&&\Big\{\Big[\hat{\vartheta}_{u}+\hat{\vartheta}_{v}+1\Big]
\Big[\hat{\vartheta}_{u}+3-{D\over2}\Big]^2\Big[\hat{\vartheta}_{u}+{5\over2}-{D\over2}\Big]
\nonumber\\
&&\hspace{0.0cm}
+{1\over u}\hat{\vartheta}_{u}\Big[\hat{\vartheta}_{u}+2-{D\over2}\Big]
\Big[\hat{\vartheta}_{u}+{1\over2}\Big]\Big[\hat{\vartheta}_{u}-\hat{\vartheta}_{v}+2
-{D\over2}\Big]\Big\}C_{_0}^{(1)}=0
\;,\nonumber\\
&&\Big\{\Big[\hat{\vartheta}_{u}+\hat{\vartheta}_{v}+1\Big]
\Big[\hat{\vartheta}_{u}-\hat{\vartheta}_{v}+2-{D\over2}\Big]
+{1\over v}\hat{\vartheta}_{v}\Big[\hat{\vartheta}_{v}
-{1\over2}\Big]\Big\}C_{_0}^{(1)}=0\;,
\label{1LoopC15}
\end{eqnarray}
where $\hat{\vartheta}_{u}u^\alpha=u^\alpha(\hat{\vartheta}_{u}+\alpha)$ is used.
Defining the auxiliary functions
\begin{eqnarray}
&&F_{_t}(u,u^\prime,v)=\Big[4\hat{\vartheta}_{u}+2\hat{\vartheta}_{u^\prime}
+6-D\Big]C_{_0}^{(2)}(u,u^\prime,v)
\nonumber\\
&&\hspace{2.1cm}=2\hat{\vartheta}_{u^\prime}C_{_0}^{(3)}(u,u^\prime,v)
\label{1LoopC16}
\end{eqnarray}
under the restriction $u=u^\prime=p_{_1}^2/p_{_2}^2$,
we present the system of PDEs satisfied by $F_{_{t}}$ as
\begin{eqnarray}
&&\Big\{\Big[\hat{\vartheta}_{u}+3-{D\over2}\Big]\Big[2\hat{\vartheta}_{u}+\hat{\vartheta}_{u^\prime}+6-D\Big]
\Big[2\hat{\vartheta}_{u}+\hat{\vartheta}_{u^\prime}+5-D\Big]
\Big[\hat{\vartheta}_{u}+\hat{\vartheta}_{v}+1\Big]
\nonumber\\
&&\hspace{0.0cm}
+{1\over u}\Big[\hat{\vartheta}_{u}+2-{D\over2}\Big]
\Big[2\hat{\vartheta}_{u}+\hat{\vartheta}_{u^\prime}+1\Big]
\Big[2\hat{\vartheta}_{u}+\hat{\vartheta}_{u^\prime}\Big]
\Big[\hat{\vartheta}_{u}-\hat{\vartheta}_{v}+2-{D\over2}\Big]\Big\}F_{_t}=0
\;,\nonumber\\
&&\Big\{\Big[2\hat{\vartheta}_{u}+\hat{\vartheta}_{u^\prime}+5-D\Big]
\Big[\hat{\vartheta}_{u^\prime}+2-{D\over2}\Big]
-{1\over u^\prime}\Big[\hat{\vartheta}_{u^\prime}-2+{D\over2}\Big]
\Big[2\hat{\vartheta}_{u}+\hat{\vartheta}_{u^\prime}
+1\Big]\Big\}F_{_t}=0
\;,\nonumber\\
&&\Big\{\Big[\hat{\vartheta}_{u}+\hat{\vartheta}_{v}+1\Big]
\Big[\hat{\vartheta}_{u}-\hat{\vartheta}_{v}+2-{D\over2}\Big]
+{1\over v}\hat{\vartheta}_{v}\Big[\hat{\vartheta}_{v}
-{1\over2}\Big]\Big\}F_{_t}=0\;.
\label{1LoopC17}
\end{eqnarray}

In the kinematic region $|p_{_1}^2|\ge\max(|p_{_2}^2|,\;|(p_{_1}-p_{_2})^2|)$,
the massless one-loop 3-point function
can be obtained by interchanging $p_{_1}\leftrightarrow p_{_2}$ in Eq.(\ref{1LoopC09-a}).
Additionally it is straightly shown that the first term $u^{D/2-3}C_{_0}^{(1)}(1/u,v)$
satisfies the system of homogeneous linear PDEs in Eq.(\ref{1LoopC15}), and the terms
$u^{D/2-3}C_{_0}^{(2)}(1/u,1/u^\prime,v)$, $u^{D/2-3}C_{_0}^{(3)}(1/u,1/u^\prime,v)$
satisfy the system of homogeneous linear PDEs in Eq.(\ref{1LoopC17}), respectively.

Using the Laurent series of $C_{_0}^{(1)}$, $C_{_0}^{(2)}$ and $C_{_0}^{(3)}$
around space-time dimensions $D=4$ in Eq.(\ref{1LoopC42}), one gets
\begin{eqnarray}
&&C_{_0}=
{i\over(4\pi)^{2}p_{_2}^2}
\sum\limits_{n=0}^\infty\sum_{r=0}^n
\Big\{{(-)^{n+r}2^r(n+r)!\over(1+2n)r!(n-r)!(2r-1)!!}
\nonumber\\
&&\hspace{1.0cm}\times
\Big[-\ln{p_{_1}^2\over p_{_2}^2}+{2\over1+2n}\Big]
\Big({p_{_1}^2\over p_{_2}^2}\Big)^n\Big({(p_{_1}\cdot p_{_2})^2\over p_{_1}^2p_{_2}^2}\Big)^r
\nonumber\\
&&\hspace{1.0cm}
-{p_{_1}\cdot p_{_2}\over p_{_2}^2}\cdot
{(-)^{n+r}2^{1+r}(1+n+r)!\over (2+2n)r!(n-r)!(2r+1)!!}
\nonumber\\
&&\hspace{1.0cm}\times
\Big[-\ln{p_{_1}^2\over p_{_2}^2}+{1\over1+n}\Big]
\Big({p_{_1}^2\over p_{_2}^2}\Big)^n\Big({(p_{_1}\cdot p_{_2})^2\over p_{_1}^2p_{_2}^2}\Big)^r\Big\}
+{\cal O}(\varepsilon)\;.
\label{1LoopC43}
\end{eqnarray}

Actually a very-well known result of the scalar integral of one-loop massless triangle
diagram is published in Ref.~\cite{Davydychev87,Davydychev92}. In order to compare
the well known result with ours explicitly, we give the Eq.(7)
of Ref.~\cite{Davydychev87} in our conventions as
\begin{eqnarray}
&&C_{_0}={-\Gamma(2-{D\over2})\Gamma({D\over2}-1)\over(4\pi)^{D/2}i^{D-1}\Gamma(D-3)
(-p_{_2}^2)^{3-D/2}}
\nonumber\\
&&\hspace{1.0cm}\times
\Big\{\Gamma({D\over2}-2)\;\varsigma^{D/2-2}F_{_4}
\left(\left.\begin{array}{cc}1,&{D\over2}-1\\
3-{D\over2},&{D\over2}-1\end{array}\right|u,\;\varsigma\right)
\nonumber\\
&&\hspace{1.0cm}
+\Gamma({D\over2}-2)\;u^{D/2-2}F_{_4}
\left(\left.\begin{array}{cc}1,&{D\over2}-1\\
{D\over2}-1,&3-{D\over2}\end{array}\right|u,\;\varsigma\right)
\nonumber\\
&&\hspace{1.0cm}
-\Gamma({D\over2}-2)\;F_{_4}
\left(\left.\begin{array}{cc}1,&3-{D\over2}\\
3-{D\over2},&3-{D\over2}\end{array}\right|u,\;\varsigma\right)
\nonumber\\
&&\hspace{1.0cm}
+\Gamma(2-{D\over2})\Gamma(D-3)\;(u\varsigma)^{D/2-2}F_{_4}
\left(\left.\begin{array}{cc}D-3,&{D\over2}-1\\
{D\over2}-1,&{D\over2}-1\end{array}\right|u,\;\varsigma\right)
\Big\}\;,
\label{1LoopC44}
\end{eqnarray}
with $k=p_{_1}+p_{_2}$, $u=p_{_1}^2/p_{_2}^2$, $\varsigma=k^2/p_{_2}^2$.
Adopting the reduction formulae of Eq.(\ref{1LoopB0-46}), and the expansion
\begin{eqnarray}
&&\;_{_2}F_{_1}\left(\left.\begin{array}{c}1,\;\;1-\varepsilon\\
1+\varepsilon\end{array}\right|x\right)
=(1-x)^{2\varepsilon-1}\Big\{1+2\varepsilon^2L_{i_2}(x)+{\cal O}(\varepsilon^3)\Big\}\;,
\label{1LoopC49}
\end{eqnarray}
one derives
\begin{eqnarray}
&&C_{_0}={i\over(4\pi)^{2}p_{_2}^2}\Phi(u,\varsigma)\;,
\label{1LoopC49-a}
\end{eqnarray}
where the concrete expression of $\Phi(u,\varsigma)$ can be found in Eq.(2.11) of Ref.~\cite{Davydychev92}.
Certainly the expression of Eq.(\ref{1LoopC49-a}) is obtained in the region
$\lambda_{_{u,\varsigma}}^2\ge0$, which is pointed explicitly in Ref.~\cite{A.I.Davydychev93}.
As $|u|\sim|\zeta|=|p_{_1}\cdot p_{_2}/p_{_2}^2|\ll1$,
$\lambda_{_{u,\varsigma}}^2=-4u+4\zeta^2<0$. The analytic continuation
to the region $\lambda_{_{u,\varsigma}}^2<0$ can be done by
the reduction formulae in Eq.(\ref{1LoopB0-46}) and the quadratic transformation in
Eq.(\ref{1LoopB0-51}):
\begin{eqnarray}
&&C_{_0}={-i\Gamma({D\over2}-1)\over(4\pi)^{D/2}\Gamma(D-3)
(-p_{_2}^2)^{3-D/2}\lambda_{_{u,\varsigma}}}
\nonumber\\
&&\hspace{1.0cm}\times
\Big\{\Gamma(\varepsilon)\Gamma(-\varepsilon)\;\varsigma^{-\varepsilon}
\Big[{\varepsilon\over\varepsilon-{1\over2}}\Big({\lambda_{_{u,\varsigma}}^2\over4u}\Big)^{1/2}
\;_{_2}F_{_1}\left(\left.\begin{array}{c}{1\over2},\;\;{1\over2}-\varepsilon\\
{3\over2}-\varepsilon\end{array}\right|
-{\lambda_{_{u,\varsigma}}^2\over4u}\right)
\nonumber\\
&&\hspace{1.0cm}
+{\Gamma(1+\varepsilon)\Gamma({1\over2}-\varepsilon)\over
\Gamma({1\over2})}\Big({\lambda_{_{u,\varsigma}}^2\over4u}\Big)^{\varepsilon}\Big]
\nonumber\\
&&\hspace{1.0cm}
+\Gamma(\varepsilon)\Gamma(-\varepsilon)\;u^{-\varepsilon}
\Big[{\varepsilon\over\varepsilon-{1\over2}}\Big({\lambda_{_{u,\varsigma}}^2\over4\varsigma}\Big)^{1/2}
\;_{_2}F_{_1}\left(\left.\begin{array}{c}{1\over2},\;\;{1\over2}-\varepsilon\\
{3\over2}-\varepsilon\end{array}\right|
-{\lambda_{_{u,\varsigma}}^2\over4\varsigma}\right)
\nonumber\\
&&\hspace{1.0cm}
+{\Gamma(1+\varepsilon)\Gamma({1\over2}-\varepsilon)\over
\Gamma({1\over2})}\Big({\lambda_{_{u,\varsigma}}^2\over4\varsigma}\Big)^{\varepsilon}\Big]
\nonumber\\
&&\hspace{1.0cm}
-\Gamma(\varepsilon)\Gamma(-\varepsilon)
\Big[{\varepsilon\over\varepsilon-{1\over2}}\Big({\lambda_{_{u,\varsigma}}^2\over4u\varsigma}\Big)^{1/2}
\;_{_2}F_{_1}\left(\left.\begin{array}{c}{1\over2},\;\;{1\over2}-\varepsilon\\
{3\over2}-\varepsilon\end{array}\right|
-{\lambda_{_{u,\varsigma}}^2\over4u\varsigma}\right)
\nonumber\\
&&\hspace{1.0cm}
+{\Gamma(1+\varepsilon)\Gamma({1\over2}-\varepsilon)\over
\Gamma({1\over2})}\Big({\lambda_{_{u,\varsigma}}^2\over4u\varsigma}\Big)^{\varepsilon}\Big]
\nonumber\\
&&\hspace{1.0cm}
+\Gamma^2(\varepsilon)\Gamma(1-2\varepsilon)\;(u\varsigma)^{-\varepsilon}
\Big[{1\over\lambda_{_{u,\varsigma}}}\Big]^{-2\varepsilon}\Big\}
\label{1LoopC53}
\end{eqnarray}
It is easy to derive the expansion when $\varepsilon\rightarrow0$
\begin{eqnarray}
&&\;_{_2}F_{_1}\left(\left.\begin{array}{c}{1\over2},\;\;{1\over2}-\varepsilon\\
{3\over2}-\varepsilon\end{array}\right|x\right)=
{1\over\sqrt{x}}\arcsin\sqrt{x}
+\varepsilon\Big[-{2\over\sqrt{x}}\arcsin\sqrt{x}
+{\ln(4x)\over\sqrt{x}}\arcsin\sqrt{x}
\nonumber\\
&&\hspace{4.3cm}
+{1\over\sqrt{x}}{\rm Cl}_{_2}(2\arcsin\sqrt{x})\Big]+\cdots\;.
\label{1LoopC54}
\end{eqnarray}
Using this expansion and some well-known relations of {\rm arcsine}
and Clausen functions, we get
\begin{eqnarray}
&&C_{_0}={i2\over(4\pi)^{2}p_{_2}^2\sqrt{-\lambda_{_{u,\varsigma}}^2}}
\nonumber\\
&&\hspace{1.0cm}\times
\Big\{\ln\Big(-{\lambda_{_{u,\varsigma}}^2\over4\zeta}\Big)\arcsin\sqrt{-{\lambda_{_{u,\varsigma}}^2\over4u}}
+{\rm Cl}_{_2}(2\arcsin\sqrt{-{\lambda_{_{u,\varsigma}}^2\over4u}})
\nonumber\\
&&\hspace{1.0cm}
+\ln\Big(-{\lambda_{_{u,\varsigma}}^2\over4\varsigma}\Big)\arcsin\sqrt{-{\lambda_{_{u,\varsigma}}^2\over4\varsigma}}
+{\rm Cl}_{_2}(2\arcsin\sqrt{-{\lambda_{_{u,\varsigma}}^2\over4\varsigma}})
\nonumber\\
&&\hspace{1.0cm}
-\ln\Big(-{\lambda_{_{u,\varsigma}}^2\over4u\varsigma}\Big)\arcsin\sqrt{-{\lambda_{_{u,\varsigma}}^2\over4u\varsigma}}
-{\rm Cl}_{_2}(2\arcsin\sqrt{-{\lambda_{_{u,\varsigma}}^2\over4u\varsigma}})
\nonumber\\
&&\hspace{1.0cm}
-\ln \varsigma\;\arcsin\sqrt{-{\lambda_{_{u,\varsigma}}^2\over4u}}
-\ln u\;\arcsin\sqrt{-{\lambda_{_{u,\varsigma}}^2\over4\varsigma}}\Big\}\;.
\label{1LoopC53a}
\end{eqnarray}
in the region $\lambda_{_{u,\varsigma}}^2<0$.
Consequently the power series of Eq.(\ref{1LoopC53a}) around $p_{_1}^2/p_{_2}^2=0$,
$p_{_1}\cdot p_{_2}/p_{_2}^2=0$ is derived as
\begin{eqnarray}
&&C_{_0}={i\over(4\pi)^{2}p_{_2}^2}
\Big\{2-\ln{p_{_1}^2\over p_{_2}^2}-\Big(1-\ln{p_{_1}^2\over p_{_2}^2}\Big)
{p_{_1}\cdot p_{_2}\over p_{_2}^2}-\Big({2\over9}-{1\over3}\ln{p_{_1}^2\over p_{_2}^2}\Big)
{p_{_1}^2\over p_{_2}^2}+\cdots\Big\}\;,
\label{1LoopC55}
\end{eqnarray}
which coincides with Eq.(\ref{1LoopC43}) exactly. In other words, the
result of Eq.(\ref{1LoopC43}) represents the double power series around $p_{_1}^2/p_{_2}^2=0$,
$p_{_1}\cdot p_{_2}/p_{_2}^2=0$ of the result from
Ref.~\cite{Davydychev87,Davydychev92} in the region $\lambda_{_{u,\varsigma}}^2<0$.

For massive one-loop triangle diagram, the corresponding scalar integral is
written as
\begin{eqnarray}
&&C_{_0}=\int{d^Dq\over(2\pi)^D}{1\over(q^2-m_{_0}^2)((q+p_{_1})^2-m_{_1}^2)
((q-p_{_2})^2-m_{_2}^2)}
\nonumber\\
&&\hspace{0.5cm}=-i
{2^3(m_{_0}^2m_{_1}^2m_{_2}^2)^{D/2-1}\over(4\pi)^{3D/2}}
\int d^D{\bf x}_{_1}d^D{\bf x}_{_2}k_{_{D/2-1}}(m_{_0}|{\bf x}_{_1}
-{\bf x}_{_2}|)
\nonumber\\
&&\hspace{1.0cm}\times
k_{_{D/2-1}}(m_{_1}x_{_1})k_{_{D/2-1}}(m_{_2}x_{_2})
\exp\{i({\bf x}_{_1}\cdot{\bf p}_{_1}+{\bf x}_{_2}\cdot{\bf p}_{_2})\}\;.
\label{1LoopC18}
\end{eqnarray}
Adopting the addition theorem from Ref.~\cite{G.N.Watson44}
\begin{eqnarray}
&&k_{_{\mu}}(|{\bf x}-{\bf x}^\prime|)=\sum\limits_{n=0}^\infty
(\mu+n)C_{_n}^\mu(\hat{\bf x}\cdot\hat{\bf x}^\prime)\Big({xx^\prime\over4}\Big)^n
\Big[i_{_{\mu+n}}(x)k_{_{\mu+n}}(x^\prime)\Theta(x^\prime-x)
\nonumber\\
&&\hspace{2.6cm}
+i_{_{\mu+n}}(x^\prime)k_{_{\mu+n}}(x)\Theta(x-x^\prime)\Big]\;,
\label{1LoopC19}
\end{eqnarray}
one presents the final result similar to Eq.(\ref{1LoopC09-a}).
Here the power series of the modified Bessel function with
imaginary argument is written as
\begin{eqnarray}
&&i_{_{\mu}}(x)=\sum\limits_{n=0}^\infty{1\over n!\Gamma(1+\mu+n)}
\Big({x\over2}\Big)^{2n}\;.
\label{1LoopC20}
\end{eqnarray}
In order to obtain the multiple hypergeometric functions in the kinematic region
$$m_{_0}^2\ge\max(|p_{_1}^2|,\;|p_{_2}^2|,\;|(p_{_1}\cdot p_{_2})^2/p_{_1}^2p_{_2}^2|,\;
m_{_1}^2,\;m_{_2}^2),$$ we also derive the indispensably radial integral for $i_{_{\mu}}(t)$ as
\begin{eqnarray}
&&\int_0^\infty dt\Big({t\over2}\Big)^{2\rho-1}i_{_\mu}(t)=
{\sin(\mu\pi-\rho\pi)\sin({\mu\pi\over2}-{\pi\over4})
\over\pi\cos(\rho\pi-{\mu\pi\over2}-{\pi\over4})}
\Gamma(\rho)\Gamma(\rho-\mu)\;.
\label{1LoopC21}
\end{eqnarray}
As far as we know, the expression of Eq.(\ref{1LoopC21}) is firstly presented here.
Inserting Eq.(\ref{1LoopC21}) into the well-known relation in Ref.~\cite{G.N.Watson44}
\begin{eqnarray}
&&k_{_\mu}(t)={\Gamma(\mu)\Gamma(1-\mu)\over2}\Big\{-\Big({t\over2}\Big)^{-2\mu}i_{_{-\mu}}(t)
+i_{_\mu}(t)\Big\}\;,
\label{1LoopC22}
\end{eqnarray}
one gets the first radial integral of Eq.(\ref{base15}) explicitly.
This provides a cross check on our result in Eq.(\ref{1LoopC21}).
The analytic expression of the scalar integral for one-loop massive triangle
diagrams contains three terms in the vicinity of each coordinate axis
of independent variables $p_{_1}^2/p_{_2}^2$, $(p_{_1}\cdot p_{_2})^2/p_{_1}^2p_{_2}^2$,
$m_{_i}^2/p_{_2}^2\;(i=0,\;1,\;2)$. Defining the auxiliary functions
similar to that in Eq.(\ref{1LoopC16}), one finds those terms satisfying two systems of
homogeneous linear PDEs similar to that presented
in Eq.(\ref{1LoopC15}) and Eq.(\ref{1LoopC17}), respectively.

The scalar integral of the one-loop box diagram can also be analyzed by
the hypergeometric functions, the corresponding analytic expression of the
scalar integral contains 27 terms in the vicinity of each coordinate axis
of independent variables. Defining several auxiliary functions, one finds
those terms satisfying three systems of homogeneous linear PDEs respectively.
It is worth noting that a well-known analysis on one-loop massless box diagram is also presented in
Ref.~\cite{Davydychev93-1}. In order to shorten the length of context,
we release our analyses in detail elsewhere.

\section{The system of linear PDEs as the stationary
condition of a functional\label{sec7}}
\indent\indent
As stated above, the $B_{_0}$ function is formulated
through the double hypergeometric functions of Eq.(\ref{1LoopB0-3a}) for the kinematic region
$\sqrt{|x|}+\sqrt{|y|}\le1$, where the function $\varphi_{_1}(x,y)$
satisfies the system of PDEs in Eq.(\ref{1LoopB0-5}). Meanwhile, the $B_{_0}$ function is
formulated through the double hypergeometric functions of Eq.(\ref{1LoopB0-7a}) for the kinematic region
$1+\sqrt{|x|}\le\sqrt{|y|}$, i.e. $\sqrt{|\xi|}+\sqrt{|\eta|}\le1$,
where the function $\varphi_{_2}(\xi,\eta)$ satisfies the  system of PDEs in Eq.(\ref{1LoopB0-9}).
Now the congruence between the systems of homogeneous linear
PDEs in Eq.(\ref{1LoopB0-5}) and Eq.(\ref{1LoopB0-9}) can be proved directly. Applying
$\varphi_{_2}(\xi,\eta)=(-y)^{2-D/2}\varphi_{_1}(x,y),\;\xi=1/y,\;
\eta=x/y$, we have
\begin{eqnarray}
&&{\partial\varphi_{_2}\over\partial\xi}=(-)^{2-D/2}\Big\{
-xy^{3-D/2}{\partial\varphi_{_1}\over\partial x}
-y^{4-D/2}{\partial\varphi_{_1}\over\partial y}
+(D/2-2)y^{3-D/2}\varphi_{_1}\Big\}
\;,\nonumber\\
&&{\partial\varphi_{_2}\over\partial\eta}=(-)^{2-D/2}
y^{3-D/2}{\partial\varphi_{_1}\over\partial x}
\;,\nonumber\\
&&{\partial^2\varphi_{_2}\over\partial\xi^2}=(-)^{2-D/2}\Big\{
x^2y^{4-D/2}{\partial^2\varphi_{_1}\over\partial x^2}
+2xy^{5-D/2}{\partial^2\varphi_{_1}\over\partial x\partial y}
+y^{6-D/2}{\partial^2\varphi_{_1}\over\partial y^2}
\nonumber\\
&&\hspace{1.6cm}
+(6-D)xy^{4-D/2}{\partial\varphi_{_1}\over\partial x}
+(6-D)y^{5-D/2}{\partial\varphi_{_1}\over\partial y}
+(D/2-2)(D/2-3)y^{4-D/2}\varphi_{_1}\Big\}
\;,\nonumber\\
&&{\partial^2\varphi_{_2}\over\partial\eta^2}=(-)^{2-D/2}
y^{4-D/2}{\partial^2\varphi_{_1}\over\partial x^2}
\;,\nonumber\\
&&{\partial^2\varphi_{_2}\over\partial\xi\partial\eta}=(-)^{2-D/2}
\Big\{-xy^{4-D/2}{\partial^2\varphi_{_1}\over\partial x^2}
-y^{5-D/2}{\partial^2\varphi_{_1}\over\partial x\partial y}
+(D/2-3)y^{4-D/2}{\partial\varphi_{_1}\over\partial x}\Big\}\;.
\label{system-1}
\end{eqnarray}
Inserting those derivatives into the first PDE of Eq.(\ref{1LoopB0-9}), one derives
\begin{eqnarray}
&&(-y)^{3-D/2}\Big\{(\hat{\vartheta}_{x}+\hat{\vartheta}_{y}+2-{D\over2})
(\hat{\vartheta}_{x}+\hat{\vartheta}_{y}+3-D)-{1\over y}\hat{\vartheta}_{y}
(\hat{\vartheta}_{y}+1-{D\over2})\Big\}\varphi_{_1}=0\;,
\label{system-2}
\end{eqnarray}
which is equal to the second  PDE of Eq.(\ref{1LoopB0-5}) exactly.
Inserting those derivatives into the second PDE of Eq.(\ref{1LoopB0-9}),
one similarly finds
\begin{eqnarray}
&&(-y)^{3-D/2}\Big\{{1\over x}\hat{\vartheta}_{x}
(\hat{\vartheta}_{x}+1-{D\over2})-{1\over y}\hat{\vartheta}_{y}
(\hat{\vartheta}_{y}+1-{D\over2})\Big\}\varphi_{_1}=0\;,
\label{system-3}
\end{eqnarray}
which is equal to the difference between two PDEs of Eq.(\ref{1LoopB0-5})
correspondingly.
\begin{figure}[h]
\setlength{\unitlength}{1cm}
\centering
\vspace{0.0cm}\hspace{-1.5cm}
\includegraphics[height=8cm,width=8.0cm]{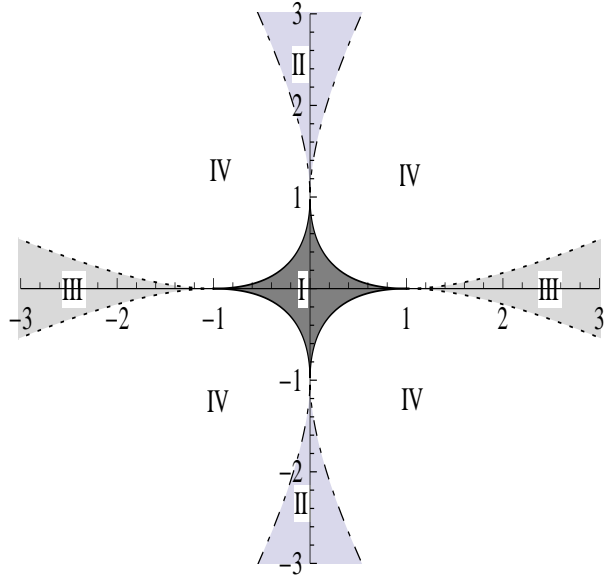}
\vspace{0cm}
\caption[]{The dark gray region I is $\sqrt{|x|}+\sqrt{|y|}\le1$ ($\sqrt{|s|}+\sqrt{|t|}\le1$),
the gray region II is $1+\sqrt{|x|}\le\sqrt{|y|}$ ($1+\sqrt{|s|}\le\sqrt{|t|}$), the light gray region III
is $1+\sqrt{|y|}\le\sqrt{|x|}$ ($1+\sqrt{|t|}\le\sqrt{|s|}$), respectively.
Where the analytic expressions in double hypergeometric
functions are given in Eq.(\ref{system-5}) (Eq.(\ref{system-10})). The continuation of corresponding
solutions to the white region IV is made through the systems of linear PDEs
in Eq.(\ref{system-8c}) (Eq.(\ref{system-13c})).}
\label{fig1}
\end{figure}
In other words, the $B_{_0}$ function can be formulated as
\begin{eqnarray}
&&B_{_0}(p^2)={i\Gamma(1+\varepsilon)\over(1-2\varepsilon)(4\pi)^2}
\Big({4\pi\mu^2\over-p^2}\Big)^\varepsilon\Phi_{_B}(x,y)\;,
\label{system-4}
\end{eqnarray}
where
\begin{eqnarray}
&&\Phi_{_B}(x,y)=\left\{\begin{array}{ll}
\varphi_{_1}(x,y)\;,&\sqrt{|x|}+\sqrt{|y|}\le1\\
(-y)^{D/2-2}\varphi_{_2}({1\over y},{x\over y})\;,&1+\sqrt{|x|}\le\sqrt{|y|}\\
(-x)^{D/2-2}\varphi_{_2}({1\over x},{y\over x})\;,&1+\sqrt{|y|}\le\sqrt{|x|}
\end{array}\right.
\label{system-5}
\end{eqnarray}
satisfies the system of homogeneous linear PDEs:
\begin{eqnarray}
&&\Big\{(\hat{\vartheta}_{x}+\hat{\vartheta}_{y}+2-{D\over2})
(\hat{\vartheta}_{x}+\hat{\vartheta}_{y}+3-D)-{1\over x}\hat{\vartheta}_{x}
(\hat{\vartheta}_{x}+1-{D\over2})\Big\}\Phi_{_B}=0
\;,\nonumber\\
&&\Big\{(\hat{\vartheta}_{x}+\hat{\vartheta}_{y}+2-{D\over2})
(\hat{\vartheta}_{x}+\hat{\vartheta}_{y}+3-D)-{1\over y}\hat{\vartheta}_{y}
(\hat{\vartheta}_{y}+1-{D\over2})\Big\}\Phi_{_B}=0\;.
\label{system-6}
\end{eqnarray}
The $B_{_0}$ function under the restriction
$y=0$ is
\begin{eqnarray}
&&\Phi_{_B}(x,0)=F_{_B}(x)=\left\{\begin{array}{ll}
\varphi_{_1}(x,0)\;,&|x|\le1\\
(-x)^{D/2-2}\varphi_{_2}({1\over x},0)\;,&|x|\ge1
\end{array}\right.\;.
\label{system-6-a}
\end{eqnarray}
Using the well-known relation of Gauss functions in Eq.(\ref{1LoopB0-51}),
one finds $\varphi_{_1}(x,0)=(-x)^{D/2-2}\varphi_{_2}({1\over x},0)$.
It indicates that $F_{_B}(x)$ is a continuously differentiable function
in the $x-$coordinate axis, and satisfies the first PDE
under the restriction $y=0$ in Eq.(\ref{system-6}).
Furthermore one can write down the analytic expressions of derivatives of any order for $F_{_B}(x)$
in the whole $x-$coordinate axis. Similarly $\Phi_{_B}(0,y)=F_{_B}(y)$
satisfies the second PDE under the restriction $x=0$ in Eq.(\ref{system-6}).
Because of the compatibility between two PDEs in Eq.(\ref{system-6})
and the uniqueness theorem of solution to the system of PDEs~\cite{M.E.Taylor12},
the continuation of $\Phi_{_B}(x,y)$ to the entire $x-y$ plane is made numerically
with its analytic expression on the whole $x-$axis and the system of PDEs in Eq.(\ref{system-6}).

By the system of PDEs of Eq.(\ref{system-6}), the continuation of $\Phi_{_B}$
from the kinematic regions I, II, and III to the kinematic region IV can be made numerically.
In order to perform the continuation of $\Phi_{_B}$ to the kinematic region IV,
we present its Laurent series around space-time dimensions $D=4$ as
\begin{eqnarray}
&&\Phi_{_B}(x,y)={\phi_{_B}^{(-1)}(x,y)\over\varepsilon}
+\phi_{_B}^{(0)}(x,y)+\sum\limits_{i=1}^\infty\varepsilon^i\phi_{_B}^{(i)}(x,y)\;.
\label{system-7}
\end{eqnarray}
Inserting $D=4-2\varepsilon$ and the above expansion into
the system of linear PDEs Eq.(\ref{system-6}),
one derives the systems of linear PDEs
satisfied by $\phi_{_B}^{(-1)}$, $\phi_{_B}^{(0)}$
and $\phi_{_B}^{(n)}\;(n=1,\;2,\;\cdots)$ respectively. In order to shorten the length of
text, we present those systems of linear PDEs
in appendix \ref{app2}.

As stated above, the analytic continuation of the $B_{_0}$ function
to the region IV can be made equivalently through the quadratic transformation:
\begin{eqnarray}
&&\phi_{_B}^{(-1)}(x,y)=1
\;,\nonumber\\
&&\phi_{_B}^{(0)}(x,y)=-{1\over2}\ln(xy)-{x-y\over2}\ln{x\over y}
-\lambda_{_{x,y}}\ln{1-x-y-\lambda_{_{x,y}}\over2\sqrt{xy}}\;.
\label{system-8-1}
\end{eqnarray}
Using those expressions, one easily verifies that $\phi_{_B}^{(-1)}(x,y)$ and $\phi_{_B}^{(0)}(x,y)$
satisfy the two systems of PDEs in Eq.(\ref{system-8a})
and Eq.(\ref{system-8b}) explicitly.

In the scalar integral from multi-loop Feynman diagrams,
the coefficient of the lowest power of $\varepsilon$
is generally a polynomial function of its independent variables. Since the sets with the
restrictions $x=0$ or $y=0$ are regular singularities of the system of PDEs in Eq.(\ref{system-6}),
the factors such as $(-x)^{\varepsilon},\;(-y)^\varepsilon$ induce the possible
imaginary corrections to $\phi_{_B}^{(n)}(x,y)\;(n\ge1)$.
Under this circumstance, the real and imaginary parts of $\phi_{_B}^{(n)}(x,y)\;(n\ge1)$
satisfy the system of PDEs in Eq.(\ref{system-8c}) separately.
This character of $\phi_{_B}^{(n)}(x,y)\;(n\ge1)$ provides
a cross check on the self-consistency of our cross-cuts in the Riemann planes.

Similarly the double hypergeometric function of the two-loop vacuum is written as
\begin{eqnarray}
&&V_{_2}={\Gamma^2(1+\varepsilon)\over2(4\pi)^4(1-\varepsilon)(1-2\varepsilon)}
\Big({4\pi\mu^2\over m_{_3}^2}\Big)^{2\varepsilon}m_{_3}^2\Phi_{_{v}}(s,t)\;,
\label{system-9}
\end{eqnarray}
where
\begin{eqnarray}
&&\Phi_{_v}(s,t)=\left\{\begin{array}{ll}
\varphi(s,t)\;,&\sqrt{|s|}+\sqrt{|t|}\le1\\
s^{3-D}\varphi({1\over s},{t\over s})\;,&1+\sqrt{|t|}\le\sqrt{|s|}\\
t^{3-D}\varphi({1\over t},{s\over t})\;,&1+\sqrt{|s|}\le\sqrt{|t|}
\end{array}\right.
\label{system-10}
\end{eqnarray}
satisfies the system of the PDEs
\begin{eqnarray}
&&\Big\{(\hat{\vartheta}_{s}+\hat{\vartheta}_{t}+2-{D\over2})
(\hat{\vartheta}_{s}+\hat{\vartheta}_{t}+3-D)-{1\over s}\hat{\vartheta}_{s}
(\hat{\vartheta}_{s}+1-{D\over2})\Big\}\Phi_{_v}=0
\;,\nonumber\\
&&\Big\{(\hat{\vartheta}_{s}+\hat{\vartheta}_{t}+2-{D\over2})
(\hat{\vartheta}_{s}+\hat{\vartheta}_{t}+3-D)-{1\over t}\hat{\vartheta}_{t}
(\hat{\vartheta}_{t}+1-{D\over2})\Big\}\Phi_{_v}=0\;.
\label{system-11}
\end{eqnarray}
The two-loop vacuum under the restriction $t=0$ is
\begin{eqnarray}
&&\Phi_{_v}(s,0)=F_{_v}(s)=\left\{\begin{array}{ll}
\varphi(s,0)\;,&|s|\le1\\
s^{3-D}\varphi({1\over s},0)\;,&|s|\ge1
\end{array}\right.\;.
\label{system-10-a}
\end{eqnarray}
Using the well-known relation of Eq.(\ref{1LoopB0-51}),
one also derives $\varphi(s,0)=s^{3-D}\varphi({1\over s},0)$.
It indicates that $F_{_v}(s)$ is a continuously differentiable function
in the $s-$coordinate axis, and satisfies the first PDE
with the constraint $t=0$ in Eq.(\ref{system-11}).
Similarly the continuation of the solution $\Phi_{_v}(s,t)$ to entire $s-t$ plane
is made through its analytic expression on the whole $s-$axis and the corresponding
PDEs in Eq.(\ref{system-11}).

In order to make the continuation of $\Phi_{_v}$ to the kinematic region IV numerically,
we give the Laurent series of two-loop vacuum around space-time
dimensions $D=4$ as
\begin{eqnarray}
&&\Phi_{_v}(x,y)={\phi_{_v}^{(-2)}(x,y)\over\varepsilon^2}
+{\phi_{_v}^{(-1)}(x,y)\over\varepsilon}
+\phi_{_v}^{(0)}(x,y)+\sum\limits_{i=1}^\infty\varepsilon^i\phi_{_v}^{(i)}(x,y)\;.
\label{system-12}
\end{eqnarray}
Thus, one derives the systems of PDEs
satisfied by $\phi_{_v}^{(-2)}$, $\phi_{_v}^{(-1)}$, $\phi_{_v}^{(0)}$
and $\phi_{_v}^{(n)}\;(n=1,\;2,\;\cdots)$ directly. In order to shorten the length of
text, we present those systems of PDEs in appendix \ref{app2}.

For the two-loop vacuum integral, the continuation of the corresponding expression
to the region IV can be made also with the quadratic transformation:
\begin{eqnarray}
&&\phi_{_v}^{(-2)}(s,t)=-1-s-t
\;,\nonumber\\
&&\phi_{_v}^{(-1)}(s,t)=2(s\ln s+t\ln t)
\;,\nonumber\\
&&\phi_{_v}^{(0)}(s,t)=-s\ln^2s-t\ln^2t+(1-s-t)\ln s\ln t-\lambda_{_{s,t}}\Phi(s,t)\;.
\label{system-13-1}
\end{eqnarray}
Using those expressions, one easily verifies that $\phi_{_v}^{(-2)}(s,t)$, $\phi_{_v}^{(-1)}(s,t)$
and $\phi_{_v}^{(0)}(s,t)$ satisfy three systems of PDEs in Eq.(\ref{system-13a}), Eq.(\ref{system-13b}),
and Eq.(\ref{system-13c}), respectively.

Generally for the scalar integrals of Feynman diagrams, the continuation
of the multiple hypergeometric functions from its convergent regions to the whole
kinematic domain can be made numerically through the systems of PDEs.
After obtaining the solutions $\phi_{_B}^{(n-2)},\;\phi_{_B}^{(n-1)}$
in the whole $x-y$ plane, we write the system of PDEs satisfied
by $F=x^{(c_{_1}-1)/2}y^{(c_{_2}-1)/2}\phi_{_B}^{(n)}$ as
\begin{eqnarray}
&&x{\partial^2F\over\partial x^2}-y{\partial^2F\over\partial y^2}
+{\partial F\over\partial x}-{\partial F\over\partial y}
-\Big[{(c_{_1}-1)^2\over4x}-{(c_{_2}-1)^2\over4y}\Big]F
\nonumber\\
&&\hspace{0.0cm}
-x^{(c_{_1}-1)/2}y^{(c_{_2}-1)/2}\Big(f_{_1}-f_{_2}\Big)=0
\;,\nonumber\\
&&x(1-2x){\partial^2F\over\partial x^2}+y(1-2y){\partial^2F\over\partial y^2}
-4xy{\partial^2F\over\partial x\partial y}
\nonumber\\
&&\hspace{0.0cm}
+\Big[1-2(3+a+b-c_{_1}-c_{_2})x\Big]{\partial F\over\partial x}
+\Big[1-2(3+a+b-c_{_1}-c_{_2})y\Big]{\partial F\over\partial y}
\nonumber\\
&&\hspace{0.0cm}
-\Big[{(c_{_1}-1)^2\over4x}+{(c_{_2}-1)^2\over4y}
+2(1+a-{c_{_1}+c_{_2}\over2})(1+b-{c_{_1}+c_{_2}\over2})\Big]F
\nonumber\\
&&\hspace{0.0cm}
-x^{(c_{_1}-1)/2}y^{(c_{_2}-1)/2}\Big(f_{_1}+f_{_2}\Big)=0\;,
\label{system-14}
\end{eqnarray}
with
\begin{eqnarray}
&&f_{_1}(x,y)=-(1-3x){\partial\phi_{_B}^{(n-1)}\over\partial x}
+3y{\partial\phi_{_B}^{(n-1)}\over\partial y}
-\phi_{_B}^{(n-1)}+2\phi_{_B}^{(n-2)}
\;,\nonumber\\
&&f_{_2}(x,y)=3x{\partial\phi_{_B}^{(n-1)}\over\partial x}
-(1-3y){\partial\phi_{_B}^{(n-1)}\over\partial y}
-\phi_{_B}^{(n-1)}+2\phi_{_B}^{(n-2)}\;,
\label{system-15}
\end{eqnarray}
and $a=c_{_1}=c_{_2}=0,\;b=-1$ for the $B_{_0}$ function.
Actually the system of PDEs can be
recognized as stationary conditions of the modified functional~\cite{R.Courant53}
\begin{eqnarray}
&&\Pi^*(F)=\Pi(F)+\int\limits_{\Omega}\chi(x,y)\Big\{
x(1-2x){\partial^2F\over\partial x^2}+y(1-2y){\partial^2F\over\partial y^2}
-4xy{\partial^2F\over\partial x\partial y}
\nonumber\\
&&\hspace{1.6cm}
+\Big[1-2(3+a+b-c_{_1}-c_{_2})x\Big]{\partial F\over\partial x}
+\Big[1-2(3+a+b-c_{_1}-c_{_2})y\Big]{\partial F\over\partial y}
\nonumber\\
&&\hspace{1.6cm}
-\Big[{(c_{_1}-1)^2\over4x}+{(c_{_2}-1)^2\over4y}
+2(1+a-{c_{_1}+c_{_2}\over2})(1+b-{c_{_1}+c_{_2}\over2})\Big]F
\nonumber\\
&&\hspace{1.6cm}
-x^{(c_{_1}-1)/2}y^{(c_{_2}-1)/2}\Big(f_{_1}+f_{_2}\Big)\Big\}dxdy\;,
\label{system-16}
\end{eqnarray}
where $\chi(x,y)$ denotes Lagrange multiplier, $\Omega$ represents the kinematic
region where the continuation of the solution is made numerically, and $\Pi(F)$ is the functional
of the first PDE in Eq.(\ref{system-14}):
\begin{eqnarray}
&&\Pi(F)=\int\limits_{\Omega}\Big\{
-{x\over2}\Big({\partial F\over\partial x}\Big)^2+{y\over2}\Big({\partial F\over\partial y}\Big)^2
-\Big[{(c_{_1}-1)^2\over8x}-{(c_{_2}-1)^2\over8y}\Big]F^2
\nonumber\\
&&\hspace{1.6cm}
-x^{(c_{_1}-1)/2}y^{(c_{_2}-1)/2}\Big(f_{_1}-f_{_2}\Big)F\Big\}dxdy\;.
\label{system-17}
\end{eqnarray}
Here the stationary condition of $\Pi(F)$ is the first PDE of Eq.(\ref{system-14}),
the stationary condition of the second term of Eq.(\ref{system-16}) is the second PDE
of Eq.(\ref{system-14}) which is recognized as a restriction of the system here.
Because of the boundary conditions $\Phi_{_B}(x,0)=F_{_B}(x)$, the continuation of
the solution to whole kinematic region is made numerically with finite element
method~\cite{X.C.Wang03} from Eq.(\ref{system-16}).

\begin{figure}[h]
\setlength{\unitlength}{1cm}
\centering
\vspace{0.0cm}\hspace{-1.5cm}
\includegraphics[height=8cm,width=8.0cm]{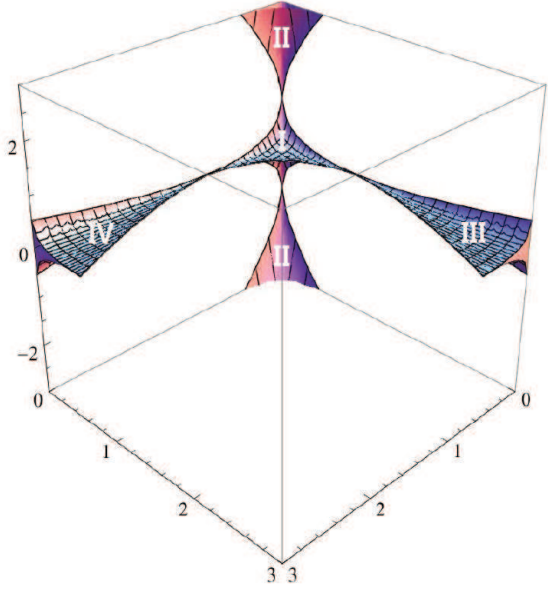}
\vspace{0cm}
\caption[]{The convergent regions of triple hypergeometric functions
in Eq.(\ref{system-19}) in the first quarter. The continuation of
the corresponding solutions to the whole kinematic domain is made numerically
through the systems of PDEs in Eq.(\ref{system-28}).}
\label{fig2}
\end{figure}

Similarly the scalar integral of two-loop sunset diagram is formulated as
\begin{eqnarray}
&&\Sigma_{_{\ominus}}(p^2)=
-{p^2\over(4\pi)^4}\Big({4\pi\mu^2\over-p^2}\Big)^{2\varepsilon}
\Gamma^2(1+\varepsilon)\Phi_{_{123}}(x_{_1},x_{_2},x_{_3})\;,
\label{system-18}
\end{eqnarray}
where
\begin{eqnarray}
&&\Phi_{_{123}}(x_{_1},x_{_2},x_{_3})=\left\{\begin{array}{cc}T_{_{123}}^{p}(x_{_1},x_{_2},x_{_3}),&
\sqrt{|x_{_1}|}+\sqrt{|x_{_2}|}+\sqrt{|x_{_3}|}\le1\\
(-x_{_3})^{D-3}T_{_{123}}^{m}({x_{_1}\over x_{_3}},{x_{_2}\over x_{_3}},{1\over x_{_3}}),&
1+\sqrt{|x_{_1}|}+\sqrt{|x_{_2}|}\le\sqrt{|x_{_3}|}\\
(-x_{_2})^{D-3}T_{_{123}}^{m}({x_{_1}\over x_{_2}},{x_{_3}\over x_{_2}},{1\over x_{_2}}),&
1+\sqrt{|x_{_1}|}+\sqrt{|x_{_3}|}\le\sqrt{|x_{_2}|}\\
(-x_{_1})^{D-3}T_{_{123}}^{m}({x_{_3}\over x_{_1}},{x_{_2}\over x_{_1}},{1\over x_{_1}}),&
1+\sqrt{|x_{_2}|}+\sqrt{|x_{_3}|}\le\sqrt{|x_{_1}|}\\
\end{array}\right.
\label{system-19}
\end{eqnarray}
satisfies the system of PDEs
\begin{eqnarray}
&&\Big\{(\sum\limits_{i=1}^3\hat{\vartheta}_{x_{_i}}+3-D)
(\sum\limits_{i=1}^3\hat{\vartheta}_{x_{_i}}+4-{3D\over2})
-{1\over x_{_1}}\hat{\vartheta}_{x_{_1}}(\hat{\vartheta}_{x_{_1}}+1
-{D\over2})\Big\}\Phi_{_{123}}=0\;,
\nonumber\\
&&\Big\{(\sum\limits_{i=1}^3\hat{\vartheta}_{x_{_i}}+3-D)
(\sum\limits_{i=1}^3\hat{\vartheta}_{x_{_i}}+4-{3D\over2})
-{1\over x_{_2}}\hat{\vartheta}_{x_{_2}}(\hat{\vartheta}_{x_{_2}}+1
-{D\over2})\Big\}\Phi_{_{123}}=0\;,
\nonumber\\
&&\Big\{(\sum\limits_{i=1}^3\hat{\vartheta}_{x_{_i}}+3-D)
(\sum\limits_{i=1}^3\hat{\vartheta}_{x_{_i}}+4-{3D\over2})
-{1\over x_{_3}}\hat{\vartheta}_{x_{_3}}(\hat{\vartheta}_{x_{_3}}+1
-{D\over2})\Big\}\Phi_{_{123}}=0\;.
\label{system-20}
\end{eqnarray}

$\Phi_{_{123}}$ under the restriction $x_{_2}=x_{_3}=0$
is given as
\begin{eqnarray}
&&\Phi_{_{123}}(x_{_1},0,0)=F_{_{123}}(x)=\left\{\begin{array}{ll}
T_{_{123}}^{p}(x_{_1},0,0)\;,&|x_{_1}|\le1\\
(-x_{_1})^{D-3}T_{_{123}}^{m}(0,0,{1\over x_{_1}})\;,&|x_{_1}|\ge1
\end{array}\right.\;.
\label{system-19-a}
\end{eqnarray}
Using the well-known relation of Gauss function in Eq.(\ref{1LoopB0-51}),
one derives $$T_{_{123}}^{p}(x_{_1},0,0)=(-x_{_1})^{D-3}T_{_{123}}^{m}(0,0,{1\over x_{_1}}).$$
The relation indicates that $F_{_{123}}(x_{_1})$ is a continuously differentiable function
of the whole $x_{_1}-$coordinate axis, and satisfies the first PDE
under the restriction $x_{_2}=x_{_3}=0$ in Eq.(\ref{system-20}).
Furthermore, one can write down the analytic expressions of derivatives of any order for $F_{_{123}}(x_{_1})$
in the whole $x_{_1}-$coordinate axis. Similarly $\Phi_{_{123}}(0,x_{_2},0)=F_{_{123}}(x_{_2})$
satisfies the second PDE under the restriction $x_{_1}=x_{_3}=0$,
and $\Phi_{_{123}}(0,0,x_{_3})=F_{_{123}}(x_{_3})$ satisfies the third PDE
under the restriction $x_{_1}=x_{_2}=0$ in Eq.(\ref{system-20}), respectively.
Because of the compatibility of the PDEs in Eq.(\ref{system-20})
and the uniqueness theorem of solution to the system of PDEs~\cite{M.E.Taylor12},
the continuation of $\Phi_{_{123}}(x_{_1},x_{_2},x_{_3})$ to
whole three dimension space of $x_{_i},\;i=1,2,3$ is made numerically
through its analytic expression on the whole $x_{_1}-$axis and the
corresponding PDEs in Eq.(\ref{system-20}).
Taking the $\Phi_{_{123}}(x_{_1},0,0)=F_{_{123}}(x_{_1})$ as boundary conditions,
one performs the continuation of $\Phi_{_{123}}$ to the entire $x_{_1}-x_{_2}$ plane
numerically through the first two homogeneous linear PDEs
under the restriction $x_{_3}=0$.
Using the solution on the whole $x_{_1}-x_{_2}$ plane as boundary conditions,
then one performs the continuation of $\Phi_{_{123}}$ to whole three dimension space
numerically by the system of PDEs in Eq.(\ref{system-20}).

In order to make the continuation of $\Phi_{_{123}}$ to whole kinematic regions numerically,
we give the Laurent series of the scalar integral from two-loop sunset around space-time
dimensions $D=4$ as
\begin{eqnarray}
&&\Phi_{_{123}}(x,y)={\phi_{_{123}}^{(-2)}(x,y)\over\varepsilon^2}
+{\phi_{_{123}}^{(-1)}(x,y)\over\varepsilon}
+\phi_{_{123}}^{(0)}(x,y)+\sum\limits_{i=1}^\infty\varepsilon^i\phi_{_{123}}^{(i)}(x,y)\;.
\label{system-25}
\end{eqnarray}
Thus one similarly derives the systems of linear PDEs
satisfied by $\phi_{_{123}}^{(-2)}$, $\phi_{_{123}}^{(-1)}$, $\phi_{_{123}}^{(0)}$
and $\phi_{_{123}}^{(n)}\;(n=1,\;2,\;\cdots)$ which are presented in appendix \ref{app2}.

Using the hypergeometric functions of Eq.(\ref{system-19}), one derives
$\phi_{_{123}}^{(-2)}=(x_{_1}+x_{_2}+x_{_3})/2$ which satisfies the system of PDEs in
Eq.(\ref{system-26}) explicitly. Since there is not the reduction formula
for the Lauricella functions, the triple hypergeometric functions of Eq.(\ref{system-19})
cannot be analytically continued outside the convergent regions. Nevertheless
the continuation of the triple hypergeometric functions of the scalar integrals from two-loop sunset diagram
to whole kinematic domain can be made numerically by the systems of PDEs.
After obtaining the solutions $\phi_{_{123}}^{(n-2)},\;\phi_{_{123}}^{(n-1)}$,
one writes the system of linear PDEs satisfied
by $F=x_{_1}^{(\gamma_{_1}-1)/2}x_{_2}^{(\gamma_{_2}-1)/2}x_{_3}^{(\gamma_{_3}-1)/2}\phi_{_{123}}^{(n)}$ as
\begin{eqnarray}
&&2x_{_1}{\partial^2F\over\partial x_{_1}^2}-x_{_2}{\partial^2F\over\partial x_{_2}^2}
-x_{_3}{\partial^2F\over\partial x_{_3}^2}
+2{\partial F\over\partial x_{_1}}-{\partial F\over\partial x_{_2}}
-{\partial F\over\partial x_{_3}}
-\Big[{(\gamma_{_1}-1)^2\over 2x_{_1}}-{(\gamma_{_2}-1)^2\over4x_{_2}}
\nonumber\\
&&\hspace{0.0cm}
-{(\gamma_{_3}-1)^2\over4x_{_3}}\Big]F
-x_{_1}^{(\gamma_{_1}-1)/2}x_{_2}^{(\gamma_{_2}-1)/2}x_{_3}^{(\gamma_{_3}-1)/2}\Big(2g_{_1}-g_{_2}-g_{_3}\Big)=0
\;,\nonumber\\
&&x_{_2}{\partial^2F\over\partial x_{_2}^2}
-x_{_3}{\partial^2F\over\partial x_{_3}^2}
+{\partial F\over\partial x_{_2}}-{\partial F\over\partial x_{_3}}
-\Big[{(\gamma_{_2}-1)^2\over4x_{_2}}
-{(\gamma_{_3}-1)^2\over4x_{_3}}\Big]F
\nonumber\\
&&\hspace{0.0cm}
-x_{_1}^{(\gamma_{_1}-1)/2}x_{_2}^{(\gamma_{_2}-1)/2}x_{_3}^{(\gamma_{_3}-1)/2}\Big(g_{_2}-g_{_3}\Big)=0
\;,\nonumber\\
&&x_{_1}(1-3x_{_1}){\partial^2F\over\partial x_{_1}^2}
+x_{_2}(1-3x_{_2}){\partial^2F\over\partial x_{_2}^2}
+x_{_3}(1-3x_{_3}){\partial^2F\over\partial x_{_2}^2}
-6x_{_1}x_{_2}{\partial^2F\over\partial x_{_1}\partial x_{_2}}
\nonumber\\
&&\hspace{0.0cm}
-6x_{_1}x_{_3}{\partial^2F\over\partial x_{_1}\partial x_{_3}}
-6x_{_2}x_{_3}{\partial^2F\over\partial x_{_2}\partial x_{_3}}
+\Big[1-3(4+\alpha+\beta-\gamma_{_1}-\gamma_{_2}-\gamma_{_3})x_{_1}\Big]
{\partial F\over\partial x_{_1}}
\nonumber\\
&&\hspace{0.0cm}
+\Big[1-3(4+\alpha+\beta-\gamma_{_1}-\gamma_{_2}-\gamma_{_3})x_{_2}\Big]
{\partial F\over\partial x_{_2}}
+\Big[1-3(4+\alpha+\beta-\gamma_{_1}-\gamma_{_2}-\gamma_{_3})x_{_3}\Big]
{\partial F\over\partial x_{_3}}
\nonumber\\
&&\hspace{0.0cm}
-\Big[\sum\limits_{i=1}^3{(\gamma_{_i}-1)^2\over4x_{_i}}
+3({3\over2}+\alpha-{\gamma_{_1}+\gamma_{_2}+\gamma_{_3}\over2})
({3\over2}+\beta-{\gamma_{_1}+\gamma_{_2}+\gamma_{_3}\over2})\Big]F
\nonumber\\
&&\hspace{0.0cm}
-x_{_1}^{(\gamma_{_1}-1)/2}x_{_2}^{(\gamma_{_2}-1)/2}
x_{_3}^{(\gamma_{_3}-1)/2}\Big(g_{_1}+g_{_2}+g_{_3}\Big)=0\;,
\label{system-29}
\end{eqnarray}
with $\alpha=-1,\;\beta=-2,\;\gamma_{_1}=\gamma_{_2}=\gamma_{_3}=0$, and
\begin{eqnarray}
&&g_{_1}(x_{_1},x_{_2},x_{_3})=-(1-5x_{_1}){\partial\phi_{_{123}}^{(n-1)}\over\partial x_{_1}}
+5x_{_2}{\partial\phi_{_{123}}^{(n-1)}\over\partial x_{_2}}
+5x_{_3}{\partial\phi_{_{123}}^{(n-1)}\over\partial x_{_3}}
-7\phi_{_{123}}^{(n-1)}+6\phi_{_{123}}^{(n-2)}
\;,\nonumber\\
&&g_{_2}(x_{_1},x_{_2},x_{_3})=5x_{_1}{\partial\phi_{_{123}}^{(n-1)}\over\partial x_{_1}}
-(1-5x_{_2}){\partial\phi_{_{123}}^{(n-1)}\over\partial x_{_2}}
+5x_{_3}{\partial\phi_{_{123}}^{(n-1)}\over\partial x_{_3}}
-7\phi_{_{123}}^{(n-1)}+6\phi_{_{123}}^{(n-2)}
\;,\nonumber\\
&&g_{_3}(x_{_1},x_{_2},x_{_3})=5x_{_1}{\partial\phi_{_{123}}^{(n-1)}\over\partial x_{_1}}
+5x_{_2}{\partial\phi_{_{123}}^{(n-1)}\over\partial x_{_2}}
-(1-5x_{_3}){\partial\phi_{_{123}}^{(n-1)}\over\partial x_{_3}}
-7\phi_{_{123}}^{(n-1)}+6\phi_{_{123}}^{(n-2)}\;,
\label{system-30}
\end{eqnarray}
for two-loop sunset diagram. In a similar way, the system of PDEs
in Eq.(\ref{system-29}) is also recognized as stationary conditions
of the modified functional
\begin{eqnarray}
&&\Pi_{_{123}}^*(F)=\Pi_{_{123}}(F)
\nonumber\\
&&\hspace{2.0cm}
+\int\limits_{\Omega}\chi_{_{23}}
\Big\{x_{_2}{\partial^2F\over\partial x_{_2}^2}
-x_{_3}{\partial^2F\over\partial x_{_3}^2}
+{\partial F\over\partial x_{_2}}-{\partial F\over\partial x_{_3}}
-\Big[{(\gamma_{_2}-1)^2\over4x_{_2}}
-{(\gamma_{_3}-1)^2\over4x_{_3}}\Big]F
\nonumber\\
&&\hspace{2.0cm}
-x_{_1}^{(\gamma_{_1}-1)/2}x_{_2}^{(\gamma_{_2}-1)/2}x_{_3}^{(\gamma_{_3}-1)/2}
\Big(g_{_2}-g_{_3}\Big)\Big\}dx_{_1}dx_{_2}dx_{_3}
\nonumber\\
&&\hspace{2.0cm}
+\int\limits_{\Omega}\chi_{_{123}}
\Big\{x_{_1}(1-3x_{_1}){\partial^2F\over\partial x_{_1}^2}
+x_{_2}(1-3x_{_2}){\partial^2F\over\partial x_{_2}^2}
+x_{_3}(1-3x_{_3}){\partial^2F\over\partial x_{_2}^2}
\nonumber\\
&&\hspace{2.0cm}
-6x_{_1}x_{_2}{\partial^2F\over\partial x_{_1}\partial x_{_2}}
-6x_{_1}x_{_3}{\partial^2F\over\partial x_{_1}\partial x_{_3}}
-6x_{_2}x_{_3}{\partial^2F\over\partial x_{_2}\partial x_{_3}}
\nonumber\\
&&\hspace{2.0cm}
+\Big[1-3(4+\alpha+\beta-\gamma_{_1}-\gamma_{_2}-\gamma_{_3})x_{_1}\Big]
{\partial F\over\partial x_{_1}}
\nonumber\\
&&\hspace{2.0cm}
+\Big[1-3(4+\alpha+\beta-\gamma_{_1}-\gamma_{_2}-\gamma_{_3})x_{_2}\Big]
{\partial F\over\partial x_{_2}}
\nonumber\\
&&\hspace{2.0cm}
+\Big[1-3(4+\alpha+\beta-\gamma_{_1}-\gamma_{_2}-\gamma_{_3})x_{_3}\Big]
{\partial F\over\partial x_{_3}}
\nonumber\\
&&\hspace{2.0cm}
-\Big[\sum\limits_{i=1}^3{(\gamma_{_i}-1)^2\over4x_{_i}}
+3({3\over2}+\alpha-{\gamma_{_1}+\gamma_{_2}+\gamma_{_3}\over2})
({3\over2}+\beta-{\gamma_{_1}+\gamma_{_2}+\gamma_{_3}\over2})\Big]F
\nonumber\\
&&\hspace{2.0cm}
-x_{_1}^{(\gamma_{_1}-1)/2}x_{_2}^{(\gamma_{_2}-1)/2}
x_{_3}^{(\gamma_{_3}-1)/2}\Big(g_{_1}+g_{_2}+g_{_3}\Big)\Big\}dx_{_1}dx_{_2}dx_{_3}\;,
\label{system-31}
\end{eqnarray}
where $\chi_{_{23}}(x_{_1},x_{_2},x_{_3}),\;\chi_{_{123}}(x_{_1},x_{_2},x_{_3})$
are Lagrange multipliers, $\Omega$ represents the kinematic
domain where the continuation of the solution is made numerically, and $\Pi_{_{123}}(F)$ is the functional
of the first PDE in Eq.(\ref{system-29}):
\begin{eqnarray}
&&\Pi_{_{123}}(F)=\int\limits_{\Omega}\Big\{
-x_{_1}\Big({\partial F\over\partial x_{_1}}\Big)^2
+{x_{_2}\over2}\Big({\partial F\over\partial x_{_2}}\Big)^2
+{x_{_3}\over2}\Big({\partial F\over\partial x_{_3}}\Big)^2
\nonumber\\
&&\hspace{2.0cm}
-\Big[{(\gamma_{_1}-1)^2\over 4x_{_1}}-{(\gamma_{_2}-1)^2\over8x_{_2}}
-{(\gamma_{_3}-1)^2\over8x_{_3}}\Big]F^2
\nonumber\\
&&\hspace{2.0cm}
-x_{_1}^{(\gamma_{_1}-1)/2}x_{_2}^{(\gamma_{_2}-1)/2}x_{_3}^{(\gamma_{_3}-1)/2}
\Big(2g_{_1}-g_{_2}-g_{_3}\Big)F\Big\}dx_{_1}dx_{_2}dx_{_3}\;.
\label{system-32}
\end{eqnarray}
Furthermore the stationary condition
of the second term of Eq.(\ref{system-31}) is the second PDEs
in Eq.(\ref{system-29}), the stationary condition of the third term of
Eq.(\ref{system-31}) is the third PDEs
in Eq.(\ref{system-29}), which are recognized as two restrictions of the system here.
Taking the expressions of corresponding functions of one coordinate axis as boundary conditions,
one performs the continuation of the solution to whole kinematic region numerically
through finite element method~\cite{X.C.Wang03}.

The expression of the scalar integral of one-loop triangle diagram
is divided into three terms. In the simplified case with three zero
virtual masses, the function $C_{_0}^{(1)}(u,v)$
is reduced as
\begin{eqnarray}
&&C_{_0}^{(1)}(u,0)={\Gamma({1\over2})\Gamma(2-{D\over2})\over2^{D-3}
\Gamma({D\over2}-{1\over2})\Gamma^2({D\over2}-1)}\left\{\begin{array}{c}
u^{D/2-2}\;_{_2}F_{_1}\left(\left.\begin{array}{c}1,\;{1\over2}\\
{D\over2}-{1\over2}\end{array}\right|-u\right),\;|u|\le1\\
u^{-1}\;_{_2}F_{_1}\left(\left.\begin{array}{c}1,\;{1\over2}\\
{D\over2}-{1\over2}\end{array}\right|-{1\over u}\right),\;|u|>1
\end{array}\right.
\label{system-33}
\end{eqnarray}
on the $u-$axis, which is continuously differentiable, and satisfies the
first PDEs under the restriction $v=0$ in Eq.(\ref{1LoopC15}).
Because of the compatibility between the PDEs in Eq.(\ref{1LoopC15})
and the uniqueness theorem of solution to the system of PDEs,
the continuation of $C_{_0}^{(1)}(u,v)$ to the entire plane of $u-v$
can be performed numerically with its analytic expression on the whole $u-$axis and the
corresponding homogeneous linear PDEs of Eq.(\ref{1LoopC15}).
For the auxiliary function $F_{_t}(u,u^\prime,v)$ relating the second and third terms,
the function is simplified as
\begin{eqnarray}
&&F_{_t}(u,u,0)={2\over\Gamma({D\over2}-1)}\sum\limits_{n=0}^\infty
\sum\limits_{q=0}^\infty{(-)^{n+q}\Gamma({D\over2}-1+n)\Gamma(3-{D\over2}+q+2n)
\over n!q!\Gamma({D\over2}+q+2n)\Gamma(D-3-q)}
\nonumber\\
&&\hspace{2.5cm}\times
\left\{\begin{array}{c}u^{n+q},\;|u|\le1\\
u^{D/2-3-q-n},\;|u|>1\end{array}\right.\;,
\label{system-34}
\end{eqnarray}
on the line $u=u^\prime,\;v=0$.
Additionally, the analytic expressions of partial derivatives of any order
for $F_{_t}$ can be given analytically under the condition $u=u^\prime$.
Through the first two PDEs of Eq.(\ref{1LoopC17}) under the restriction $v=0$,
the continuation of $F_{_t}$ from the line $u=u^\prime$ to the entire plane
of $u-u^\prime$ is performed numerically at first. With the boundary condition
$F_{_t}(u,u^\prime,0)$, the continuation of $F_{_t}$ to the whole three
dimension space is made numerically through the PDEs of Eq.(\ref{1LoopC17})
because of the compatibility of three PDEs  and the uniqueness theorem of
solution to the system of PDEs. Certainly the final solution should be
imposed on the restriction $u=u^\prime=p_{_1}^2/p_{_2}^2$.
In actual calculation, one certainly provides
the Laurent series of the scalar integrals from one-loop
triangle diagram around space-time dimensions $D=4$ at first, then numerically
performs the continuation of $C_{_0}$ to whole kinematic regions with finite element method
after recognizing the relevant PDEs as stationary
conditions of the modified functionals.

Noting that the continuation of the multiple hypergeometric functions
to whole kinematic domain can also be made numerically through the finite difference method
where the partial derivatives are approximated by finite differences
in corresponding PDEs. In principle the analytic continuation of
the convergent multiple hypergeometric functions can be performed
through multiple power series of the independent
kinematic variables, nevertheless the process is cumbersome when the
system of PDEs contains too much independent variables.

\section{Summary\label{sec8}}
\indent\indent
The equivalency between Feynman parameterization and the hypergeometric function method
can be proved by the integral representations of modified Bessel functions.
Based on the power series of Bessel functions
and some well-known formulae, the multiple hypergeometric functions
of the scalar integrals from concerned Feynman diagrams can be derived. Thus
the systems of linear homogeneous PDEs satisfied by the
scalar integrals can be established in the whole kinematic domain.
Recognizing the corresponding
system of linear PDEs as stationary conditions
of a functional under the given restrictions, we can perform the continuation
of the hypergeometric functions of scalar integrals from the convergent regions
to the whole kinematic domain through numerical methods.
For this purpose, the finite element method may be applied.
Since there are some well-known reduction formulae for the double hypergeometric series
of the $B_{_0}$ function and two-loop vacuum integral in textbook,
we take examples of the $B_{_0}$ function and two-loop vacuum integral
to elucidate the technique in detail. In addition, we also discuss
the systems of linear PDEs satisfied by the scalar integrals of two-loop sunset
and one-loop triangle diagrams briefly. In principle, this hypergeometric function method
can be used to evaluate scalar integrals from any Feynman diagrams.
We will apply this technique to evaluate the scalar
integrals from multi-loop diagrams elsewhere in the near future.

\begin{acknowledgments}
\indent\indent
The work has been supported partly by the National Natural
Science Foundation of China (NNSFC) with Grant No. 11275243,
No. 11147601, No. 11675239, No. 11535002, and No. 11705045.
\end{acknowledgments}

\appendix

\section{The Laurent series for one-loop massless $C_{_0}$ function\label{app1}}
\indent\indent
In this appendix, we present the Laurent series for one-loop massless $C_{_0}$ function
around space-time dimensions $D=4$
\begin{eqnarray}
&&\Gamma({1\over2})C_{_0}^{(1)}(u,\;v)=
{1\over\varepsilon}\sum\limits_{n=0}^\infty\sum_{r=0}^n
\Big\{{(-)^{n+r}2^r(n+r)!\over(1+2n)r!(n-r)!(2r-1)!!}u^nv^r
\nonumber\\
&&\hspace{3.3cm}
-\sqrt{uv}\cdot{(-)^{n+r}2^{1+r}(1+n+r)!\over (2+2n)r!(n-r)!(2r+1)!!}
u^nv^r\Big\}
\nonumber\\
&&\hspace{3.3cm}
+\sum\limits_{n=0}^\infty\sum_{r=0}^n
\Big\{{(-)^{n+r}2^r(n+r)!\over(1+2n)r!(n-r)!(2r-1)!!}
\nonumber\\
&&\hspace{3.3cm}\times
\Big[-2\gamma_{_{\rm E}}-\ln u+2\psi(2+2n)-\psi(1+n+r)\Big]u^nv^r
\nonumber\\
&&\hspace{3.3cm}
-\sqrt{uv}\cdot{(-)^{n+r}2^{1+r}(1+n+r)!\over (2+2n)r!(n-r)!(2r+1)!!}
\nonumber\\
&&\hspace{3.3cm}\times
\Big[-2\gamma_{_{\rm E}}-\ln u+2\psi(3+2n)-\psi(2+n+r)\Big]
u^nv^r\Big\}+\cdots
\;,\nonumber\\
&&\Gamma({1\over2})C_{_0}^{(2)}(u,\;u,\;v)=
\sum\limits_{n=0}^\infty\sum_{r=0}^n
\Big\{{(-)^{n+r}2^r(n+r)!\over (2n+1)^2r!(n-r)!(2r-1)!!}u^nv^r
\nonumber\\
&&\hspace{3.8cm}
-\sqrt{uv}\cdot
{(-)^{n+r}2^{1+r}(1+n+r)!\over(2n+2)^2r!(n-r)!(2r+1)!!}
u^nv^r\Big\}+\cdots
\;,\nonumber\\
&&\Gamma({1\over2})C_{_0}^{(3)}(u,\;u,\;v)=
-{1\over\varepsilon}\sum\limits_{n=0}^\infty\sum_{r=0}^n
\Big\{{(-)^{n+r}2^r(n+r)!\over(1+2n)r!(n-r)!(2r-1)!!}u^nv^r
\nonumber\\
&&\hspace{3.8cm}
-\sqrt{uv}\cdot
{(-)^{n+r}2^{1+r}(1+n+r)!\over (2+2n)r!(n-r)!(2r+1)!!}u^nv^r\Big\}
\nonumber\\
&&\hspace{3.8cm}
-\sum\limits_{n=0}^\infty\sum_{r=0}^n
\Big\{{(-)^{n+r}2^r(n+r)!\over(1+2n)r!(n-r)!(2r-1)!!}
\nonumber\\
&&\hspace{3.8cm}\times
\Big[{1\over1+2n}-2\gamma_{_{\rm E}}+2\psi(1+2n)-\psi(1+n+r)\Big]u^nv^r
\nonumber\\
&&\hspace{3.8cm}
-\sqrt{uv}\cdot
{(-)^{n+r}2^{1+r}(1+n+r)!\over (2+2n)r!(n-r)!(2r+1)!!}
\nonumber\\
&&\hspace{3.8cm}\times
\Big[{1\over2+2n}-2\gamma_{_{\rm E}}+2\psi(2+2n)-\psi(2+n+r)\Big]
u^nv^r\Big\}+\cdots\;.
\label{1LoopC42}
\end{eqnarray}

\section{The system of linear PDEs for
Laurent expansion around $D=4$\label{app2}}
\indent\indent
Here we present firstly the systems of linear PDEs
satisfied by $\phi_{_B}^{(-1)}$, $\phi_{_B}^{(0)}$
and $\phi_{_B}^{(n)}$ respectively as
\begin{eqnarray}
&&x(1-x){\partial^2\phi_{_B}^{(-1)}\over\partial x^2}-y^2{\partial^2\phi_{_B}^{(-1)}\over\partial y^2}
-2xy{\partial^2\phi_{_B}^{(-1)}\over\partial x\partial y}=0
\;,\nonumber\\
&&y(1-y){\partial^2\phi_{_B}^{(-1)}\over\partial y^2}-x^2{\partial^2\phi_{_B}^{(-1)}\over\partial x^2}
-2xy{\partial^2\phi_{_B}^{(-1)}\over\partial x\partial y}=0\;,
\label{system-8a}
\end{eqnarray}

\begin{eqnarray}
&&x(1-x){\partial^2\phi_{_B}^{(0)}\over\partial x^2}
-y^2{\partial^2\phi_{_B}^{(0)}\over\partial y^2}
-2xy{\partial^2\phi_{_B}^{(0)}\over\partial x\partial y}
\nonumber\\
&&+(1-3x){\partial\phi_{_B}^{(-1)}\over\partial x}
-3y{\partial\phi_{_B}^{(-1)}\over\partial y}+\phi_{_B}^{(-1)}=0
\;,\nonumber\\
&&y(1-y){\partial^2\phi_{_B}^{(0)}\over\partial y^2}
-x^2{\partial^2\phi_{_B}^{(0)}\over\partial x^2}
-2xy{\partial^2\phi_{_B}^{(0)}\over\partial x\partial y}
\nonumber\\
&&-3x{\partial\phi_{_B}^{(-1)}\over\partial x}
+(1-3y){\partial\phi_{_B}^{(-1)}\over\partial y}+\phi_{_B}^{(-1)}=0\;,
\label{system-8b}
\end{eqnarray}
$$\;\;\;\;\;\;\;\;\;\;\;\;\cdots\;\;\;\cdots\;,$$
\begin{eqnarray}
&&x(1-x){\partial^2\phi_{_B}^{(n)}\over\partial x^2}
-y^2{\partial^2\phi_{_B}^{(n)}\over\partial y^2}
-2xy{\partial^2\phi_{_B}^{(n)}\over\partial x\partial y}
\nonumber\\
&&+(1-3x){\partial\phi_{_B}^{(n-1)}\over\partial x}
-3y{\partial\phi_{_B}^{(n-1)}\over\partial y}
+\phi_{_B}^{(n-1)}-2\phi_{_B}^{(n-2)}=0
\;,\nonumber\\
&&y(1-y){\partial^2\phi_{_B}^{(n)}\over\partial y^2}
-x^2{\partial^2\phi_{_B}^{(n)}\over\partial x^2}
-2xy{\partial^2\phi_{_B}^{(n)}\over\partial x\partial y}
\nonumber\\
&&-3x{\partial\phi_{_B}^{(n-1)}\over\partial x}
+(1-3y){\partial\phi_{_B}^{(n-1)}\over\partial y}
+\phi_{_B}^{(n-1)}-2\phi_{_B}^{(n-2)}=0\;.
\label{system-8c}
\end{eqnarray}
$$\;\;\;\;\;\;\;\;\;\;\;\;\cdots\;\;\;\cdots\;.$$

Similarly the systems of linear PDEs
satisfied by $\phi_{_v}^{(-2)}$, $\phi_{_v}^{(-1)}$, $\phi_{_v}^{(0)}$
and $\phi_{_v}^{(n)}$ are:
\begin{eqnarray}
&&s(1-s){\partial^2\phi_{_v}^{(-2)}\over\partial s^2}
-t^2{\partial^2\phi_{_v}^{(-2)}\over\partial t^2}
-2st{\partial^2\phi_{_v}^{(-2)}\over\partial s\partial t}=0
\;,\nonumber\\
&&t(1-t){\partial^2\phi_{_v}^{(-2)}\over\partial t^2}
-s^2{\partial^2\phi_{_v}^{(-2)}\over\partial s^2}
-2st{\partial^2\phi_{_v}^{(-2)}\over\partial s\partial t}=0\;,
\label{system-13a}
\end{eqnarray}

\begin{eqnarray}
&&s(1-s){\partial^2\phi_{_v}^{(-1)}\over\partial s^2}
-t^2{\partial^2\phi_{_v}^{(-1)}\over\partial t^2}
-2st{\partial^2\phi_{_v}^{(-1)}\over\partial s\partial t}
\nonumber\\
&&+(1-3s){\partial\phi_{_v}^{(-2)}\over\partial s}
-3t{\partial\phi_{_v}^{(-2)}\over\partial t}+\phi_{_v}^{(-2)}=0
\;,\nonumber\\
&&t(1-t){\partial^2\phi_{_v}^{(-1)}\over\partial t^2}
-s^2{\partial^2\phi_{_v}^{(-1)}\over\partial s^2}
-2st{\partial^2\phi_{_v}^{(-1)}\over\partial s\partial t}
\nonumber\\
&&-3s{\partial\phi_{_v}^{(-2)}\over\partial s}
+(1-3t){\partial\phi_{_v}^{(-2)}\over\partial t}+\phi_{_v}^{(-2)}=0\;,
\label{system-13b}
\end{eqnarray}
$$\;\;\;\;\;\;\;\;\;\;\;\;\cdots\;\;\;\cdots\;,$$
\begin{eqnarray}
&&s(1-s){\partial^2\phi_{_v}^{(n)}\over\partial s^2}
-t^2{\partial^2\phi_{_v}^{(n)}\over\partial t^2}
-2st{\partial^2\phi_{_v}^{(n)}\over\partial s\partial t}
\nonumber\\
&&+(1-3s){\partial\phi_{_v}^{(n-1)}\over\partial s}
-3t{\partial\phi_{_v}^{(n-1)}\over\partial t}
+\phi_{_v}^{(n-1)}-2\phi_{_v}^{(n-2)}=0
\;,\nonumber\\
&&t(1-t){\partial^2\phi_{_v}^{(n)}\over\partial t^2}
-s^2{\partial^2\phi_{_v}^{(n)}\over\partial s^2}
-2st{\partial^2\phi_{_v}^{(n)}\over\partial s\partial t}
\nonumber\\
&&-3s{\partial\phi_{_v}^{(n-1)}\over\partial s}
+(1-3t){\partial\phi_{_v}^{(n-1)}\over\partial t}
+\phi_{_v}^{(n-1)}-2\phi_{_v}^{(n-2)}=0\;,
\label{system-13c}
\end{eqnarray}
$$\;\;\;\;\;\;\;\;\;\;\;\;\cdots\;\;\;\cdots\;.$$

Correspondingly we present the systems of linear PDEs
satisfied by $\phi_{_{123}}^{(-2)}$, $\phi_{_{123}}^{(-1)}$, $\phi_{_{123}}^{(0)}$
and $\phi_{_{123}}^{(n)}\;(n=1,\;2,\;\cdots)$:
\begin{eqnarray}
&&\Big\{(\sum\limits_{i=1}^3\hat{\vartheta}_{x_{_i}}-1)
(\sum\limits_{i=1}^3\hat{\vartheta}_{x_{_i}}-2)
-{1\over x_{_1}}\hat{\vartheta}_{x_{_1}}(\hat{\vartheta}_{x_{_1}}-1
\Big\}\phi_{_{123}}^{(-2)}=0\;,
\nonumber\\
&&\Big\{(\sum\limits_{i=1}^3\hat{\vartheta}_{x_{_i}}-1)
(\sum\limits_{i=1}^3\hat{\vartheta}_{x_{_i}}-2)
-{1\over x_{_2}}\hat{\vartheta}_{x_{_2}}(\hat{\vartheta}_{x_{_2}}-1
\Big\}\phi_{_{123}}^{(-2)}=0\;,
\nonumber\\
&&\Big\{(\sum\limits_{i=1}^3\hat{\vartheta}_{x_{_i}}-1)
(\sum\limits_{i=1}^3\hat{\vartheta}_{x_{_i}}-2)
-{1\over x_{_3}}\hat{\vartheta}_{x_{_3}}(\hat{\vartheta}_{x_{_3}}-1
\Big\}\phi_{_{123}}^{(-2)}=0\;,
\label{system-26}
\end{eqnarray}

\begin{eqnarray}
&&\Big\{(\sum\limits_{i=1}^3\hat{\vartheta}_{x_{_i}}-1)
(\sum\limits_{i=1}^3\hat{\vartheta}_{x_{_i}}-2)
-{1\over x_{_1}}\hat{\vartheta}_{x_{_1}}(\hat{\vartheta}_{x_{_1}}-1
\Big\}\phi_{_{123}}^{(-1)}
\nonumber\\
&&\hspace{0.0cm}
+\Big\{{1\over x_{_1}}\hat{\vartheta}_{x_{_1}}
-5\sum\limits_{i=1}^3\hat{\vartheta}_{x_{_i}}+7\Big\}\phi_{_{123}}^{(-2)}=0\;,
\nonumber\\
&&\Big\{(\sum\limits_{i=1}^3\hat{\vartheta}_{x_{_i}}-1)
(\sum\limits_{i=1}^3\hat{\vartheta}_{x_{_i}}-2)
-{1\over x_{_2}}\hat{\vartheta}_{x_{_2}}(\hat{\vartheta}_{x_{_2}}-1
\Big\}\phi_{_{123}}^{(-1)}
\nonumber\\
&&\hspace{0.0cm}
+\Big\{{1\over x_{_2}}\hat{\vartheta}_{x_{_2}}
-5\sum\limits_{i=1}^3\hat{\vartheta}_{x_{_i}}+7\Big\}\phi_{_{123}}^{(-2)}=0\;,
\nonumber\\
&&\Big\{(\sum\limits_{i=1}^3\hat{\vartheta}_{x_{_i}}-1)
(\sum\limits_{i=1}^3\hat{\vartheta}_{x_{_i}}-2)
-{1\over x_{_3}}\hat{\vartheta}_{x_{_3}}(\hat{\vartheta}_{x_{_3}}-1
\Big\}\phi_{_{123}}^{(-1)}
\nonumber\\
&&\hspace{0.0cm}
+\Big\{{1\over x_{_3}}\hat{\vartheta}_{x_{_3}}
-5\sum\limits_{i=1}^3\hat{\vartheta}_{x_{_i}}+7\Big\}\phi_{_{123}}^{(-2)}=0\;,
\label{system-27}
\end{eqnarray}
$$\cdots\;\;\cdots\;\;\cdots\;\;\cdots\;,$$
\begin{eqnarray}
&&\Big\{(\sum\limits_{i=1}^3\hat{\vartheta}_{x_{_i}}-1)
(\sum\limits_{i=1}^3\hat{\vartheta}_{x_{_i}}-2)
-{1\over x_{_1}}\hat{\vartheta}_{x_{_1}}(\hat{\vartheta}_{x_{_1}}-1
\Big\}\phi_{_{123}}^{(n)}
\nonumber\\
&&\hspace{0.0cm}
+\Big\{{1\over x_{_1}}\hat{\vartheta}_{x_{_1}}
-5\sum\limits_{i=1}^3\hat{\vartheta}_{x_{_i}}+7\Big\}\phi_{_{123}}^{(n-1)}
-6\phi_{_{123}}^{(n-2)}=0\;,
\nonumber\\
&&\Big\{(\sum\limits_{i=1}^3\hat{\vartheta}_{x_{_i}}-1)
(\sum\limits_{i=1}^3\hat{\vartheta}_{x_{_i}}-2)
-{1\over x_{_2}}\hat{\vartheta}_{x_{_2}}(\hat{\vartheta}_{x_{_2}}-1
\Big\}\phi_{_{123}}^{(n)}
\nonumber\\
&&\hspace{0.0cm}
+\Big\{{1\over x_{_2}}\hat{\vartheta}_{x_{_2}}
-5\sum\limits_{i=1}^3\hat{\vartheta}_{x_{_i}}+7\Big\}\phi_{_{123}}^{(n-1)}
-6\phi_{_{123}}^{(n-2)}=0\;,
\nonumber\\
&&\Big\{(\sum\limits_{i=1}^3\hat{\vartheta}_{x_{_i}}-1)
(\sum\limits_{i=1}^3\hat{\vartheta}_{x_{_i}}-2)
-{1\over x_{_3}}\hat{\vartheta}_{x_{_3}}(\hat{\vartheta}_{x_{_3}}-1
\Big\}\phi_{_{123}}^{(n)}
\nonumber\\
&&\hspace{0.0cm}
+\Big\{{1\over x_{_3}}\hat{\vartheta}_{x_{_3}}
-5\sum\limits_{i=1}^3\hat{\vartheta}_{x_{_i}}+7\Big\}\phi_{_{123}}^{(n-1)}
-6\phi_{_{123}}^{(n-2)}=0\;,
\label{system-28}
\end{eqnarray}
$$\cdots\;\;\cdots\;\;\cdots\;\;\cdots\;.$$


\begin{thebibliography}{99}
\bibitem{CMS}{\rm CMS} Collaboration, Phys.~Lett.~B{\bf 716}(2012)30.
\bibitem{ATLAS}{\rm ATLAS} Collaboration, Phys.~Lett.~B{\bf 716}(2012)1.
\bibitem{PDG}K.~A.~Olive {\it et al}.(Particle Data Group), Chin.~Phys.~C,{\bf38}(2014)090001.

\bibitem{K.G.Chetyrkin81}K.~G.~Chetyrkin, F.~V.~Tkachov, Nucl.~Phys.~B{\bf 192}(1981)159.
\bibitem{G.'t.Hooft79}Gerard't~Hooft, M.~J.~G.~Veltman, Nucl.~Phys.~B{\bf 153}(1979)365.
\bibitem{A.Denner11}A.~Denner, S.~Dittmaier, Nucl.~Phys.~B{\bf 844}(2011)199.
\bibitem{V.A.Smirnov12}V.~A.~Smirnov,{\it Analytic Tools for Feynman Integrals},
(Springer, Heidelberg 2012), and references therein.
\bibitem{R.J.Gonsalves83}R.~J.~Gonsalves, Phys.~Rev.~D{\bf 28}(1983)1542.
\bibitem{V.A.Smirnov99}V.~A.~Smirnov, Phys.~Lett.~B{\bf 460}(1999)397.
\bibitem{J.B.Tausk99}V.~A.~Smirnov, Phys.~Lett.~B{\bf 469}(1999)225.

\bibitem{A.V.Kotikov91-1}A.~V.~Kotikov,  Phys.~Lett.~B{\bf 254}(1991)158.
\bibitem{A.V.Kotikov91-2}A.~V.~Kotikov,  Phys.~Lett.~B{\bf 259}(1991)314.
\bibitem{A.V.Kotikov91-3}A.~V.~Kotikov,  Phys.~Lett.~B{\bf 267}(1991)123.
\bibitem{A.V.Kotikov91-4}A.~V.~Kotikov,  Mod.~Phys.~Lett.~A{\bf 6}(1991)677.
\bibitem{A.V.Kotikov92-1}A.~V.~Kotikov,  Int.~J.~Mod.~Phys.~A{\bf 7}(1992)1977.
\bibitem{S.Laporta96}S.~Laporta, E.~Remiddi, Phys.~Lett.~B{\bf 379}(1996)283.
\bibitem{S.Laporta97}S.~Laporta, E.~Remiddi, Acta.~Phys.~Polon~B{\bf 28}(1997)959.
\bibitem{E.Remiddi97}E.~Remiddi, Nuovo~Cim.~A{\bf110}(1997)1435.
\bibitem{S.Laporta00}S.~Laporta, Int.~J.~Mod.~Phys.~A{\bf 15}(2000)5087.
\bibitem{K.Melnikov00-1}K.~Melnikov, T.~van~Ritbergen, Phys.~Lett.~B{\bf 482}(2000)99.
\bibitem{K.Melnikov00-2}K.~Melnikov, T.~van~Ritbergen, Nucl.~Phys.~B{\bf 591}(2000)515.

\bibitem{M.Y.Kalmykov10}V.~V.~Bytev, M.~Y.~Kalmykov, and B.~A.~Kniehl,
Nucl.~Phys.~B{\bf 836}(2010)129.
\bibitem{M.Y.Kalmykov12}M.~Y.~Kalmykov, and B.~A.~Kniehl,
Phys.~Lett.~B{\bf 714}(2012)103.
\bibitem{M.Y.Kalmykov13}V.~V.~Bytev, M.~Y.~Kalmykov, and B.~A.~Kniehl,
Comput.~Phys.~Commun.~{\bf184}(2013)2332.
\bibitem{M.Y.Kalmykov17}M.~Y.~Kalmykov, and B.~A.~Kniehl,
JHEP{\bf 1707}(2017)031.

\bibitem{R.N.Lee10-1}R.~N.~Lee, Nucl.~Phys.~B{\bf 830}(2010)474.
\bibitem{R.N.Lee10-2}R.~N.~Lee, A.~V.~Smirnov, V.~A.~Smirnov, JHEP{\bf 1004}(2010)020.
\bibitem{R.N.Lee11-1}R.~N.~Lee, A.~V.~Smirnov, V.~A.~Smirnov, Eur.~Phys.~J.~C{\bf71}(2011)1708.
\bibitem{R.N.Lee11-2}R.~N.~Lee, A.~V.~Smirnov, V.~A.~Smirnov, JHEP{\bf 1004}(2010)020.
\bibitem{R.N.Lee11-3}R.~N.~Lee, I.~S.~Terekhov, JHEP{\bf 1101}(2011)068.
\bibitem{R.N.Lee12-1}R.~N.~Lee, A.~V.~Smirnov, V.~A.~Smirnov, Nucl.~Phys.~B{\bf 856}(2012)95.
\bibitem{R.N.Lee12-2}R.~N.~Lee, V.~A.~Smirnov, JHEP{\bf 1212}(2012)104.

\bibitem{V.A.Smirnov02}V.~A.~Smirnov,{\it Applied Asymptotic Expansions
in Momenta and Masses} (Springer, Heidelberg 2002), and references therein.

\bibitem{K.Hepp66}K.~Hepp, Commun.~Math.~Phys.~{\bf2}(1966)301.
\bibitem{E.R.Speer77}E.~R.~Speer, Ann.~Inst.~H.~Poincar\'{e}~{\bf23}(1977)1.
\bibitem{T.Kaneko10}T.~Kaneko, T.~Ueda, Comput.~Phys.~Common.~{\bf181}(2010)1352.

\bibitem{M.E.Taylor12}M.~E.~Taylor, {\it Partial differential equations}
(Springer, Heidelberg 2012).

\bibitem{E.Mendels78}E.~Mendels, Nuo.~Cim.~A{\bf45}(1978)87.
\bibitem{F.A.Berends94}F.~A.~Berends, M.~B\"{o}hm, M.~Buza, R.~Scharf,
Z.~Phys.~C{\bf63}(1994)227.
\bibitem{K.G.Chetyrkin83}K.~G.~Chetyrkin, S.~G.~Gorishnii, S.~A.~Larin,
F.~V.~Tkachov,  Phys.~Lett.~B{\bf 132}(1983)351.
\bibitem{K.G.Chetyrkin80}K.~G.~Chetyrkin, A.~L.~Kataev,
F.~V.~Tkachov, Nucl.~Phys.~B{\bf 174}(1980)345.

\bibitem{G.N.Watson44}G.~N.~Watson, {\it A Treatise on the Theory of Bessel Functions}
(Cambridge University Press 1944).


\bibitem{L.J.Slater66}L.~J.~Slater, {\it Generalised Hypergeometric Functions}
(Cambridge University Press 1966).

\bibitem{A.Denner93}Seeing, for example, Eq.(5.16) in
D.~Bardin, G.~Passarino, {\it The Standard Model in the Making} (Clarendon Press 1999).
\bibitem{A.I.Davydychev93}A.~I.~Davydychev, J.~B.~Tausk,
Nucl.~Phys.~B{\bf 397}(1993)123.
\bibitem{Davydychev87}E.~E.~Boos, and A.~I.~Davydychev, Vestn.~Mosk.~Univ.~{\bf28}(1987)8.
\bibitem{Davydychev92}A.~I.~Davydychev, J.~Phys.~A{\bf25}(1992)5587.
\bibitem{Davydychev93-1}A.~I.~Davydychev, Phys.~Lett.~B{\bf 305}(1993)136.
\bibitem{R.Courant53}R.~Courant, D.~Hilbert, {\it Methods of mathematical physics}
(Interscience Publishers 1953).
\bibitem{X.C.Wang03}X.~C.~Wang, {\it Finite element method}
(Tsinghua University Press 2003, in Chinese).
\bibitem{H.Bateman53}H.~Bateman, and A.~Erdelyi, {\it Higher transcendental Functions},
McGraw-Hill, New York, 1953.
\end{thebibliography}
\end{document}